\documentclass[aps, 10pt, prd,
               notitlepage, twocolumn, superscriptaddress,
               longbibliography,
               nofootinbib, floatfix]{revtex4-2}

\usepackage{epsfig}
\usepackage{url}
\usepackage[linktocpage,breaklinks]{hyperref}
\usepackage[usenames,dvipsnames]{xcolor}

\usepackage[normalem]{ulem}
\usepackage{microtype}

\usepackage{latexsym}
\usepackage{epsfig}
\usepackage{amsmath}
\usepackage{mathtools}
\usepackage{amssymb}
\usepackage{wasysym}
\usepackage{graphicx}
\usepackage{dcolumn}
\usepackage{verbatim}
\usepackage{enumerate,mdwlist}
\usepackage[titletoc]{appendix}
\usepackage{amsfonts}
\usepackage{fancyvrb}
\usepackage{tikz}
\usepackage{braket}
\usetikzlibrary{calc}
\usepackage[export]{adjustbox}
\usepackage{bm}

\usepackage[normalem]{ulem}

\apptocmd{\sloppy}{\hbadness 10000\relax}{}{}

\newcommand{\ba}{\begin{eqnarray}}
\newcommand{\ea}{\end{eqnarray}}
\newcommand{\be}{\begin{equation}}
\newcommand{\ee}{\end{equation}}

\newcommand{\nn}{\nonumber}

\definecolor{grey}{rgb}{0.4,0.4,0.4}
\definecolor{dullmagenta}{rgb}{0.4,0,0.4}
\definecolor{darkblue}{rgb}{0,0,0.4}
\definecolor{midblue}{rgb}{0,0,0.5}
\definecolor{midred}{rgb}{0.5,0,0}
\definecolor{orange}{rgb}{1,0.5,0}
\definecolor{lightbrown}{rgb}{0.75,0.5,0.25}
\definecolor{tan}{cmyk}{0.14,0.42,0.56,0}
\definecolor{djunglegreen}{cmyk}{0.99,0,0.52,0}
\definecolor{lightgreen}{rgb}{0,1,0}
\definecolor{olivegreen}{cmyk}{0.64,0,0.95,0.40}
\definecolor{midgreen}{rgb}{0.0,0.675,0.0}
\definecolor{darkgreen}{rgb}{0,0.5,0}
\definecolor{ceruleanblue}{rgb}{0.0, 0.2, 0.7}
\definecolor{burgundy}{rgb}{0.5, 0.0, 0.13}
\definecolor{hvred}{RGB}{186,12,47}

\hypersetup{
    colorlinks=true,
    linkcolor=ceruleanblue,
    filecolor=ceruleanblue,
    urlcolor=midblue,
    citecolor=burgundy,
}

\newcommand{\aei}{\affiliation{Max Planck Institute for Gravitational Physics (Albert Einstein Institute), D-14476 Potsdam, Germany}}
\newcommand{\fub}{\affiliation{Fachbereich Physik, Freie Universit\"at Berlin, D-14195 Berlin, Germany}}
\newcommand{\uiuc}{\affiliation{Department of Physics and Illinois Center for Advanced Studies of the Universe,\\University of Illinois Urbana-Champaign, Urbana, Illinois 61801, USA}}

\usepackage[all]{xy}
\usepackage{amsfonts}

\makeatletter

\def\l@subsubsection#1#2{}
\makeatother

\usepackage{hhline}
\newcommand\abs[1]{\left\lvert#1\right\rvert}

\begin{document}

\title{Eikonal gravitational-wave lensing in Einstein-aether theory}

\author{Julius Streibert}           \fub \aei
\author{Hector O. Silva}            \aei \uiuc
\author{Miguel Zumalac\'arregui}    \aei

\begin{abstract}
Einstein-aether theory provides a model to test the validity of local Lorentz invariance in gravitational interactions. The speed of gravitational waves as measured from the binary neutron star event GW170817 sets stringent limits on Einstein-aether theory, but only on a combination of the theory's
free parameters. For this reason, a significant part of the theory's parameter space remains unconstrained by observations.
Motivated by this, we explore the propagation of gravitational waves in Einstein-aether theory over an inhomogeneous background (i.e., gravitational wave lensing) as a potential mechanism to break the degeneracies
between the theory's free parameters, and hence enable new constraints on the theory to be obtained.
We reduce our analysis to gravitational waves that pass far from the lens' center and short wavelength signals, both compared to the lens' gravitational radius (eikonal limit).
By applying these approximations and bringing the field equations into the form of the so-called kinetic matrix and applying a formalism known as the propagation eigenstate framework, we find that the speed of gravitational waves is modified by inhomogeneities in the aether field. However, the modification is common to both gravitational polarizations and vanishes in the limit in which gravitational waves propagate with luminal speed. This lens-dependent gravitational wave speed contrasts with the lens-induced birefringence observed in other theories beyond general relativity, like Horndeski's theory. While the potential to improve tests based on gravitational-wave speed is limited, our formalism sets the basis to fully describe signal propagation over inhomogeneous spacetimes in Einstein-aether theory and other extensions of general relativity.
\end{abstract}

\maketitle

\section{Introduction}

Theories of gravity other than Einstein's general relativity (GR), termed “beyond GR”, are motivated, among other reasons, by unexplained phenomena like dark matter and dark energy as well as by attempts to construct a theory of quantum gravity~\cite{Berti:2015itd}.
Some theories beyond GR can be categorized by which types of fundamental fields they admit in addition to the metric tensor. Of these, theories with additional scalar fields, such as Horndeski's theory, are among the most well studied~\cite{Clifton:2011jh,Berti:2015itd}. Other theories of modified gravity have received comparatively less attention~\cite{Shankaranarayanan:2022wbx}. This includes vector-tensor theories.

The introduction of a vector field to a theory can break local Lorentz invariance (LLI) by introducing a preferred direction at each point in spacetime~\cite{Jacobson:2000xp}. Such violations of LLI are of theoretical interest. In particular, the motivation for investigating theories violating LLI stems from theories of quantum gravity, some of which utilize or even require a violation of LLI~\cite{Mattingly:2005re}. LLI violations are expected if spacetime is discrete at the Planck scale, as predicted by certain realizations of quantum gravity~\cite{Oost:2018tcv}. Note that LLI can only be tested up to the maximum energy and velocity scale that is experimentally accessible: the possibility of violations at even higher energies remains~\cite{Jacobson:2000xp}.

Einstein-aether theory is a vector-tensor theory that violates LLI by means of an additional unit timelike vector field, the aether. The theory was originally conceived by
Jacobson and
Mattingly as “gravity with a dynamical preferred frame”~\cite{Jacobson:2000xp} and is well-posed in all viable regions of its parameter space~\cite{Sarbach:2019yso}. The theory admits spherically symmetric solutions~\cite{Eling:2003rd} including analogs to stars~\cite{Eling:2006df} and black holes~\cite{Eling:2006ec}. By making some additional assumptions Einstein-aether theory becomes the IR limit of an extended form of Hořava-Lifshitz theory~\cite{Jacobson:2010mx}, linking it to the quantum gravity motivations for violations of LLI.

A wide variety of both theoretical and observational constraints can be imposed on Einstein-aether theory. While some of its parameters (for example, the Eddington-Robertson-Schiff parameters) match those of GR~\cite{Eling:2003rd}, others deviate. For example, Einstein-aether theory admits two forms of gravitational constant, local and cosmological, both of which can in principle deviate from the GR case~\cite{Carroll:2004ai}. Other parameters that can differ from GR are the post-Newtonian preferred frame parameters both in the weak field~\cite{Graesser:2005bg,Foster:2005dk} and in the strong field~\cite{Yagi:2013ava}.

Each of the five fluctuations propagates with a different speed, the value of which depends on the theory's coupling constants~\cite{Jacobson:2004ts}.
The presence of additional polarizations can be probed by gravitational interferometers~\cite{LIGOScientific:2020tif,LIGOScientific:2021sio,Takeda:2021hgo,Takeda:2023mhl}, although the coincident arrival of signals with different helicity is not expected, due to differences in speed accumulating over the signal's travel time.

The (necessarily positive) mode energy densities of Einstein-aether theory were first calculated by Eling
~\cite{Eling:2005zq}. The same year it was shown that Einstein-aether mode speeds cannot be significantly subluminal by argument of
gravitational Cherenkov radiation~\cite{Elliott:2005va}. Radiation damping at leading post-Newtonian order due to Einstein-aether radiation was first discussed in Ref.~\cite{Foster:2006az}, and this opened the door to constraints derived from observations of binary pulsars~\cite{Foster:2007gr,Yagi:2013ava}. The derivation of equations of motion and gravitational waveforms at higher post-Newtonian orders, by means of post-Minkowskian theory, is currently underway~\cite{Taherasghari:2023rwn,Taherasghari:2025mlf}.

Stringent limits on Einstein-aether can be obtained by comparing the arrival times of gravitational and electromagnetic radiation emitted by the same source~\cite{Ezquiaga:2017ekz,Creminelli:2017sry,Baker:2017hug,Sakstein:2017xjx,Lombriser:2015sxa,Bettoni:2016mij}. This was possible after the observation of a neutron-star merger, GW170817~\cite{LIGOScientific:2017vwq}, the prompt emission of a short gamma-ray burst~\cite{LIGOScientific:2017zic}, and other electromagnetic counterparts~\cite{LIGOScientific:2017ync}. GW170817 was shown to be consistent with only standard gravitational wave (GW) polarizations~\cite{LIGOScientific:2018dkp,Takeda:2020tjj}: therefore, the short time delay between gravitational and electromagnetic signals at $\sim 40$\,Mpc translated into an exquisite limit on the speed of “standard” GWs. The theory parameters that affect this quantity were thus exquisitely constrained~\cite{Gong:2018cgj,Oost:2018tcv}.

GW170817 shifted the landscape of constraints on Einstein-aether theory. Since tensor wave speeds were constrained to a precision of $10^{-15}$, previously used series expansion became invalid~\cite{Oost:2018tcv}. In particular, this made old bounds obtained from binary pulsar observations obsolete and allowed new ones to be derived~\cite{Gupta:2021vdj}. There have also been attempts to constrain Einstein-aether theory with waveform modeling and parameter estimation based on both GW170817 and GW190425~\cite{Schumacher:2023cxh}.

All tests of GW propagation in Einstein-aether theory have assumed a perfectly homogeneous background. While a reasonable first step to model GW propagation on the average universe~\cite{Ezquiaga:2018btd}, this approach neglects inhomogeneities in the metric and Einstein-aether field. Inhomogeneities in the metric are responsible for gravitational lensing, the deflection, magnification, and delay of signals propagating through the universe. In some cases, gravitational lensing can split a source into multiple images and profoundly distort their shape. Among many applications, gravitational lensing of electromagnetic sources has provided tests of gravity on extragalactic and cosmological scales~\cite{Reyes:2010tr,Collett:2018gpf}. Observations of lensed GWs can be used in a similar way~\cite{Mukherjee:2019wcg} and will also improve our ability to test the presence of additional polarizations~\cite{Goyal:2020bkm,MaganaHernandez:2022ayv}, modified speeds~\cite{Collett:2016dey,Fan:2016swi}, and modified propagation~\cite{Finke:2021znb,Narola:2023viz,Takeda:2024ghe}. In addition, it can be used to constrain the mass of gravitons~\cite{Chung:2021rcu} and the charged hair of black holes~\cite{Deka:2024ecp}.

Besides lensing, novel propagation effects can be caused by inhomogeneities in the additional fields, e.g., the Einstein-aether. A nonhomogeneous configuration allows interactions between excitations with different helicity (scalar, vector, and tensor fluctuations). In addition to modifying the propagation speed, nonsymmetric configurations can introduce birefringence, a speed difference between polarizations with the same helicity. Lens-induced GW birefringence happens in scalar-tensor theories, leading to different propagation speed for the $+$ and $\times$ polarization~\cite{Ezquiaga:2020dao}. Beyond-Einstein birefringence can be tested without the need of an electromagnetic counterpart, producing limits comparable to those of the neutron-star merger~\cite{Goyal:2023uvm}. Other classes of theories are predict birefringence between the left- and right-polarized GWs~\cite{Zhao:2019xmm,Wang:2021gqm,Jenks:2023pmk,Lagos:2024boe}.

The purpose of this paper is to examine the role that inhomogeneities in the aether field can have on the propagation of GWs, a phenomenon we will generically refer to as ``GW lensing.''\footnote{
Specifically, our calculations generalize the gravitational time delay and deflection. These results can be used to derive a modified lens equation, which potentially depends on the signal's polarization and frequency~\cite{Ezquiaga:2020dao,Goyal:2023uvm}. The modified lens equation can then be used to describe known lensing phenomena (e.g.~multiple images, distortion of extended sources) and new effects (e.g.~birefringence, dispersion~\cite{Menadeo:2024uoq}).
}
Our goal is to establish how the propagation of standard and additional polarizations depends on the theory parameters and whether they can lead to novel tests, analogous to GW birefringence. Applications to specific lens systems will be left for subsequent work.

To study GW lensing in Einstein-aether theory we will first review the basics of the theory itself: we discuss Einstein-aether theory and its GWs without lensing and with a flat Minkowski background in
Sec.~\ref{sec:aetheory}. We introduce lensing by switching to a non-Minkowski background in
Sec.~\ref{sec:gw_lensing}. We utilize certain approximations and initially obtain the new propagating modes to the lowest order in these approximations. In
Sec.~\ref{sec:higher_orders}, we then turn to consecutively higher orders. Finally we take stock of our results and of what future steps can be taken in
Sec.~\ref{sec:summary_outlook}.

We use the mostly plus metric signature and use geometrical units with $c = G = 1$, unless stated otherwise.

\section{Einstein-Aether theory}\label{sec:aetheory}

In this section, we review the basics of Einstein-aether theory. In Sec.~\ref{subsec:action_field_eq}, we present the theory's action and field equations. We then linearize the theory around a Minkowski background and
identify the theory's degrees of freedom in Sec.~\ref{subsec:linearized_theory_dof}. Finally, we review the current observational
constraints on the theory in Sec.~\ref{subsec:theory_constraints}.

\subsection{Action and field equations}\label{subsec:action_field_eq}

Einstein-aether theory
is the most general vector-tensor theory of gravity that satisfies the following conditions~\cite{Will:1993hxu,Jacobson:2000xp}:
\begin{enumerate}
\item The vector field $u^\alpha$, called the aether, is constrained to be unit timelike, $u^\alpha u_\alpha=-1$.
\item The equations of motion are of at most second order in derivatives.
\item The equations of motion are of at most second order in the aether.
\end{enumerate}

The aether is introduced in order to break LLI by introducing a preferred direction at each point in spacetime~\cite{Jacobson:2000xp}. Since a static aether would violate general covariance, the aether must instead be dynamic~\cite{Jacobson:2000xp}, itself subject to field equations. A dynamic aether with an unconstrained modulus typically evolves toward zero, returning the GR limit and restoring LLI. For this reason, we demand the aether to be unit timelike.

We now construct the theory's action. In principle, it can contain Lagrangians for gravity, aether, and matter fields, as well as gravity-aether and matter-aether couplings.
Here, we will assume that aether and matter do not couple directly. This leaves us with the Einstein-Hilbert term for gravity, $R$, the standard matter Lagrangian $\mathcal{L}_{\rm m}$ as well as a new Einstein-aether Lagrangian density $\mathcal{L}_{\text{ae}}$ containing the gravity-aether coupling:
\begin{equation}
S = \frac{1}{2\kappa_{\text{ae}}} \int {\rm d}^4 x \, \sqrt{-g}
\,
\left( R + \mathcal{L}_{\text{ae}}+2\kappa_{\text{ae}} \mathcal{L}_m\right).
\label{eq:action}
\end{equation}
Here, $\kappa_{\text{ae}}$ is the Einstein-aether equivalent of the Einstein gravitational constant, which may in principle differ from the GR value of $8\pi$.\footnote{To be precise, $\kappa_{\text{ae}}$ can be shown to be related to two of the four coupling constants of Einstein-aether theory, $c_1$ and $c_4$: $\kappa_{\text{ae}}=4\pi (2-c_1-c_4)$~\cite{Jacobson:2007veq}.} The Einstein-aether Lagrangian density contains two terms:
\begin{align}
\mathcal{L}_{\text{ae}} &= -M^{\alpha\beta}_{\phantom{\alpha\beta}\mu\nu} u^\mu_{;\alpha} u^\nu_{;\beta} +\lambda \left(g_{\alpha\beta} u^\alpha u^\beta +1\right).
\end{align}
The first contains the possible gravity-aether couplings while the second enforces the
unit timelike constraint for the aether via a Lagrange multiplier.

The tensor $M^{\alpha\beta}_{\phantom{\alpha\beta}\mu\nu}$ contains the four dimensionless coupling constants, $c_1$ through $c_4$, of Einstein-aether theory:
\begin{align}
M^{\alpha\beta}_{\phantom{\alpha\beta}\mu\nu} &= c_1 g^{\alpha\beta} g_{\mu\nu} + c_2 \delta^\alpha_\mu \delta^\beta_\nu + c_3 \delta^\alpha_\nu \delta^\beta_\mu - c_4 u^\alpha u^\beta g_{\mu\nu}.
\end{align}
We also use the following notation for linear combinations of the coupling constants:
\begin{align}
	c_{ijk\dots}&= c_i+c_j+c_k+\dots,\\
    c_+&= c_1+c_3,\label{eq:cplus}\\
    c_-&= c_1-c_3.\label{eq:cminus}
\end{align}
General relativity is recovered when all coupling constants, $c_1$ through $c_4$, vanish.

The theory's field equations follow from varying Einstein-aether action with respect to the Lagrange multiplier, the metric tensor and the aether. The Lagrange multiplier variation enforces the constraint:
\begin{align}
	g_{\alpha\beta} u^\alpha u^\beta &= -1.
\end{align}
The metric variation gives us the Einstein-aether equivalent of the Einstein field equations:
\begin{align}
	G^{\alpha\beta} - T^{\alpha\beta}_{\text{ae}} &= \kappa_{\text{ae}} T_m^{\alpha\beta},
\end{align}
where
$T^{\alpha\beta}_{\text{ae}}$ is called the aether stress-energy tensor. It can be defined in terms of the aether current $J^{\alpha}_{\phantom\alpha\mu}$ and the aether acceleration $a^\mu$:
\begin{align}
T^{\alpha\beta}_{\text{ae}}&=
-D_\mu \left[
u^{(\beta} J^{\alpha)\mu }
- J^{\mu(\alpha} u^{\beta)}
- J^{(\alpha\beta)} u^\mu
\right]\nn
\\&\mathrel{\phantom{=}}
-c_1 \left[
\left(D_\mu u^\alpha\right) \left(D^\mu u^\beta\right)
-\left(D^\alpha u_\mu\right) \left(D^\beta u^\mu\right)
\right]\nn
\\&\mathrel{\phantom{=}}
+ c_4 a^\alpha a^\beta
+ \lambda  u^\alpha u^\beta
- \frac{1}{2} g^{\alpha\beta} J^{\delta}_{\phantom\delta\sigma} D_\delta u^\sigma,\label{eq:aether_stress_energy_tensor}
\\
J^{\alpha}_{\phantom\alpha\mu}&= M^{\alpha\beta}_{\phantom{\alpha\beta}\mu\nu} D_\beta u^\nu,\\
a^\mu &= u^\alpha D_\alpha u^\mu.
\end{align}

The variation with respect to the aether yields another set of field equations:
\begin{align}
\text{\AE}_\mu&= D_\alpha J^{\alpha}_{\phantom\alpha\mu}+ c_4  a_\alpha D_\mu u^\alpha+\lambda u_\mu = 0.
\end{align}

At this stage, $\lambda$ is usually immediately eliminated~\cite{Eling:2003rd,Foster:2006az,Oost:2018tcv}. For this, it will prove convenient to split the four components of the aether field equation across two new equations by multiplying with the aether and taking symmetric and antisymmetric parts:
\begin{alignat}{2}
0&=\text{\AE}^{(\alpha} u^{\beta)}&&=u^{(\beta} D_\mu J^{\mu|\alpha)}+ c_4 a_\mu u^{(\beta} D^{\alpha)} u^\mu+\lambda u^\alpha u^\beta,\label{eq:ae_symmetric}\\
0&=\text{\AE}^{[\alpha} u^{\beta]}&&=u^{[\beta} D_\mu J^{\mu|\alpha]}+ c_4 a_\mu u^{[\beta} D^{\alpha]} u^\mu.
\end{alignat}
Solving Eq.~\eqref{eq:ae_symmetric} for $D_\mu \left(J^{\mu(\alpha}u^{\beta)}\right)$ and substituting the result in Eq.~\eqref{eq:aether_stress_energy_tensor}, we find:
\begin{align}
	T^{\alpha\beta}_{\text{ae}}&=
	-D_\mu \left[
	u^{(\beta} J^{\alpha)\mu }
	- J^{(\alpha\beta)} u^\mu
	\right]\nn
    \\&\mathrel{\phantom{=}}
	- \, c_1 \left[
	\left(D_\mu u^\alpha\right) \left(D^\mu u^\beta\right)
	-\left(D^\alpha u_\mu\right) \left(D^\beta u^\mu\right)
	\right]\nn
	\\&\mathrel{\phantom{=}}
	+ \, c_4 a^\alpha a^\beta
	+ J^{\mu(\alpha} D_\mu u^{\beta)}
	-c_4 a_\mu u^{(\beta} D^{\alpha)} u^\mu\nn
 \\&\mathrel{\phantom{=}}
	- \frac{1}{2} g^{\alpha\beta} J^{\delta}_{\phantom\delta\sigma} D_\delta u^\sigma.
\end{align}

$\text{\AE}_\mu$ has four components, and one variable was eliminated by using the symmetric part of its field equation. It then follows that the antisymmetric part has three remaining components. We choose the three entries $\text{\AE}^{[0} u^{1]}$, $\text{\AE}^{[0} u^{2]}$, and $\text{\AE}^{[0} u^{3]}$ as these independent components. All remaining entries of $\text{\AE}^{[\alpha} u^{\beta]}$ can be expressed as a linear combination of the $\text{\AE}^{[0} u^{i]}$. For example, $\text{\AE}^{[1} u^{2]}$ is equal to $u^1 \text{\AE}^{[0} u^{2]}/u^0-u^2\text{\AE}^{[0} u^{1]}/u^0$.
Our remaining set of field equations then is
\begin{align}
g_{\alpha\beta} u^\alpha u^\beta &= -1,\label{eq:unit_timelike_constraint}\\
G^{\alpha\beta} - T^{\alpha\beta}_{\text{ae}} &= \kappa_{\text{ae}} T_m^{\alpha\beta},\label{eq:eom_g}\\
\text{\AE}^{[0} u^{i]} &= 0.\label{eq:eom_aether}
\end{align}

\subsection{Linearized theory and propagating degrees of freedom}\label{subsec:linearized_theory_dof}

We now linearize our field equations by expanding the metric tensor around the flat Minkowski background, up to first order in a perturbation $h_{\mu\nu}$,
\begin{align}
	\overline{g}_{\mu\nu} &= \eta_{\mu\nu} + h_{\mu\nu},
\end{align}
and we denote linearization with an overline. By solving the background field equations it can be shown that the aether corresponding to a flat Minkowski background is $\delta^\mu_0$ \cite{Jacobson:2004ts}. Calling the aether perturbation $w^\mu$, the aether is then linearized as follows:
\begin{align}
	\overline{u}^\mu &= \delta^\mu_0 + w^\mu.
\end{align}

Note that using covariant linearization is also possible~\cite{Oost:2018tcv}. This means we could also have set $\overline{u}_\mu^{(0)} = \delta_\mu^0$, where the superscript $(0)$ denotes zeroth order in the perturbation.

We now go to combined first order in $h_{\mu\nu}$ and $w^\mu$ in the field equations. The linearization of the constraint equation~\eqref{eq:unit_timelike_constraint}
immediately eliminates one component:
\begin{align}
	0
	=
	\overline{g_{\alpha\beta} u^\alpha u^\beta} + 1
	=
	h_{00}-2 w^0 \Leftrightarrow
	w^0
	=
	\frac{h_{00}}{2}.
\end{align}

To linearize the metric field equation~\eqref{eq:eom_g} and the asymmetric aether field equation~\eqref{eq:eom_aether}
we need the linearization of $D_\beta u^{\nu}$. We note that Christoffel symbols of the Minkowski background in Cartesian coordinates have no zeroth order in the perturbation:
\begin{align}
	\Gamma^{\nu(0)}_{\beta\mu}&=0.
\end{align}

Since the partial derivative of a constant is zero, $\left(D_\beta u^{\nu}\right)^{(0)}$ vanishes. Quantities derived from covariant derivatives of the aether, such as $a^{\mu}$ and $J^{\alpha}_{\phantom\alpha\mu}$, also vanish at zeroth order.
Therefore, we see that covariant derivatives of the aether, aether acceleration, and aether current must be of first order at minimum, which means that products of any of these
are of higher order
, and any term multiplied with them must be of zeroth order.

In addition to linearizing gravity, we are right now interested in the propagation of GWs, not their emission. Therefore, we consider free GWs, set $T^{\alpha\beta}_{\rm m}=0$ and
work with the unsourced, linearized field equations:
\begin{align}
\overline{G}^{\alpha\beta} - \overline{T}^{\alpha\beta}_{\text{ae}} &= 0,\\
\overline{\text{\AE}^{[0} u^{i]}} &= 0.
\end{align}
To linear order in perturbations, the quantities that appear in the field equations are
\begin{align}
\overline{T}^{\alpha\beta}_{\text{ae}}&=
\partial_0 \overline{J}^{(\alpha\beta)}
- \delta^{(\beta}_0 \partial_\mu \overline{J}^{\alpha)\mu },\\
\overline{\text{\AE}^{[0} u^{i]}}&=-\frac{1}{2} \partial_\mu \overline{J}^{\mu i},\\
\overline{J}^{\alpha}_{\phantom{\alpha}\mu} &= M^{\alpha\beta(0)}_{\phantom{\alpha\beta}\mu\nu} \overline{D_\beta u^\nu},\\
M^{\alpha\beta(0)}_{\phantom{\alpha\beta}\mu\nu} &= c_1 \eta^{\alpha\beta} \eta_{\mu\nu} + c_2 \delta^\alpha_\mu \delta^\beta_\nu + c_3 \delta^\alpha_\nu \delta^\beta_\mu - c_4 \delta^\alpha_0 \delta^\beta_0 \eta_{\mu\nu}.
\end{align}

Instead of solving the ten components of the metric field equations and the remaining three components of the aether field equations, we will combine the two into an equation for what we call the effective tensor $X^{\alpha\beta}$:
\begin{align}
X^{\alpha\beta}&=0,\label{eq:effective_tensor_field_equations}\\
X^{00}&=\overline{G}^{00} - \overline{T}^{00}_{\text{ae}},\label{eq:X00}\\
X^{0i}&=\overline{G}^{0i} - \overline{T}^{0i}_{\text{ae}} + \overline{\text{\AE}^{[0} u^{i]}},\\
X^{i0}&=\overline{G}^{i0} - \overline{T}^{i0}_{\text{ae}} - \overline{\text{\AE}^{[i} u^{0]}},\\
X^{ij}&=\overline{G}^{ij} - \overline{T}^{ij}_{\text{ae}}.\label{eq:Xij}
\end{align}
This approach mostly matches that of Ref.~\cite{Foster:2006az}, and the key difference is discussed in Appendix~\ref{app:effective_tensor_subtleties}. Since the metric field equation contains purely symmetric terms, we do not lose any information by combining it with the equation for the purely antisymmetric $\overline{\text{\AE}^{[0} u^{i]}}$.

To solve the linearized field equations, we perform a scalar-vector-tensor (SVT) decomposition of metric and aether perturbations. Specifically, we use a decomposition introduced by Foster~\cite{Foster:2006az}:
\begin{align}
	h_{0i} &= \gamma_i + \gamma_{,i},\label{eq:SVT1}\\
	h_{ij} &= \phi_{ij} + \frac{1}{2} P_{ij} f + 2\phi_{(i,j)} + \phi_{,ij},\\
    w_i &= v_i + v_{,i}.\label{eq:SVT3}
\end{align}
where $P_{ij}$ is the following differential operator:
\begin{align}
	P_{ij} x &= \delta_{ij} \Delta x-x_{,ij}.
\end{align}
and $\Delta = \partial^{k} \partial_{k}$ is the Laplacian operator.

In Eqs.~\eqref{eq:SVT1} through~\eqref{eq:SVT3}, $\gamma$, $f$, $\phi$, and $v$ are scalars;
 $\gamma_i$, $\phi_i$, and $v_i$ are transverse vectors, meaning that their divergence vanishes, whereas $\phi_{ij}$ is a symmetric
transverse traceless matrix, meaning that $\phi^{ij}_{\phantom{ij},j}=\phi^{ij}_{\phantom{ij},i}=\phi^i_{\phantom{i}i}=0$. In essence, we decompose the vector parts into transverse and longitudinal parts and the tensor parts into transverse and longitudinal as well as traceful and traceless parts.

Note that we can freely raise and lower spatial indices since after linearization, the raising and lowering operations are done with $\eta_{\alpha\beta}$, the spatial part of which is $\delta_{ij}$. This means that from now on we also apply the Einstein sum convention to terms like $x_{kk}$.
We can easily calculate the trace $h=\eta^{\mu\nu} h_{\mu\nu}$ in terms of our decomposition:
\begin{align}
	h&=F + \Delta \phi - h_{00}.
\end{align}

Here we adopt the notation $F = \Delta f$ used by Foster~\cite{Foster:2006az} since $F$ will at some points turn out to be a more intuitive variable than $f$.

The field equations can be simplified by eliminating some components. Since $h_{\mu\nu}$ is a symmetric 4$\times$4 tensor, it has ten independent components. In terms of our decomposition, there are four scalar components $h_{00}$, $\gamma$, $f$ and $\phi$.
As transverse vectors, $\gamma_i$ and $\phi_i$ have two independent components each, as does the symmetric transverse traceless tensor $\phi_{ij}$ (three components are constrained by transverseness and one by tracelessness). We can remove some of these degrees of freedom (DOF) by using gauge invariance. Einstein-aether theory is invariant under general coordinate transformations $\varepsilon^\mu$:
\begin{align}
	x^{\prime\mu} &= x^\mu + \varepsilon^\mu.
\end{align}
Because $\varepsilon_\mu$ has four entries, we can eliminate four components with this gauge invariance.
We will choose our gauge so that
\begin{equation}
    \gamma=\phi_i=v=0\,.
\end{equation}
We show that this choice is possible in Appendix~\ref{app:gauge_considerations}.

We can now calculate the linearized covariant derivatives $\overline{D_\beta u^\nu}$, linearized aether current $\overline{J}^\alpha_{\phantom\alpha\mu}$, and, finally, the effective tensor $X^{\alpha\beta}$ in terms of the SVT decomposition of the perturbations. We find
\begingroup
\allowdisplaybreaks
\begin{align}
X_{00}
&=
-\frac{1}{2} \Delta F + \frac{c_{14}}{2} \Delta h_{00},
\\[5mm]
X_{0i}
&=
- \frac{1}{2} \Delta \gamma_i
- \frac{1}{2} \dot{F}_{,i}
-c_{14} \left(\ddot{v}_i+\ddot{\gamma}_i\right)
\nn\\&\mathrel{\phantom{=}}
+\frac{c_-}{2}\Delta \left(v_i+\gamma_i\right)
+\frac{c_{14}}{2}\dot{h}_{00,i},
\\[5mm]
X_{i0}
&=
- \frac{1}{2} \Delta \gamma_i
- \frac{1}{2} \dot{F}_{,i}
\nn\\&\mathrel{\phantom{=}}-
\frac{1}{2} \Delta \left\{
c_+ v_i
+ \partial_i \left[
\left(c_++c_2\right) \dot\phi+c_2 \dot{f}
\right]
\right\},
\\[5mm]
X_{ij}
&=
- \frac{1}{2} \left(\Delta\phi_{ij}-\ddot\phi_{ij}\right)
- \dot{\gamma}_{(i,j)}
\nn\\&\mathrel{\phantom{=}}+\frac{1}{4}P_{ij} \left(
F - \ddot{f} - 2 h_{00}  - 2 \ddot\phi
\right)
-\frac{1}{2} \ddot{f}_{,ij}
\nn\\&\mathrel{\phantom{=}}
-\frac{c_+}{2} \ddot\phi_{ij}
-
c_+ \dot{v}_{(i,j)}
\nn\\&\mathrel{\phantom{=}}-
\frac{1}{2} P_{ij} \left[c_2 \left(\ddot\phi+\ddot{f}\right)+\frac{c_+}{2}\ddot{f}\right]
\nn\\&\mathrel{\phantom{=}}-
\frac{1}{2} \partial_i \partial_j \left[\left(c_2 +c_+\right) \ddot\phi+c_2 \ddot{f}\right],
\end{align}
\endgroup
where an overdot denotes a derivative with respect to time, and $c_{\pm}$ was defined in Eqs.~\eqref{eq:cplus} and \eqref{eq:cminus}.

In addition to the SVT decomposition of the perturbations we now also perform a SVT decomposition of the effective tensor $X^{\alpha\beta}$ itself. By setting the result to zero we arrive at the new form of the field equations.
We also drop prefactors and global derivatives since we want no homogeneous or static terms in the solutions. We
obtain
four redundant components and four constraints:
\begin{align}
h_{00} &= \frac{F}{c_{14}},&
v_i &= -\frac{\gamma_i}{c_+},&
\phi &= -\frac{1+c_2}{c_{123}} f.
\end{align}
Substituting these constraints into the remaining components yields the gravitational-wave equations of Einstein-aether theory.
We obtain five wave equations for the single scalar, two vector, and two tensorial DOF, namely,
\begin{align}
0&=\Delta f - c_S^{-2} \, \ddot{f},\\
0&=\Delta \gamma_i-{c_V^{-2}} \, \ddot\gamma_i,\\
0&=\Delta\phi_{ij}-{c_T^{-2}} \, \ddot\phi_{ij},
\end{align}
which propagate with the speeds
\begin{align}
\frac{1}{c_S^2}&=\frac{(1-c_+)(2+3c_2+c_+)c_{14}}{c_{123}(2-c_{14})},\label{eq:cS}\\
\frac{1}{c_V^2}&=\frac{2c_{14}(1-c_+)}{2c_1-c_+c_-},\label{eq:cV}\\
\frac{1}{c_T^2}&=1-c_+,\label{eq:cT}
\end{align}
respectively.

During this derivation we have divided by certain combinations of the theory constants: $c_{14}$, $c_+$, and $c_{123}$ for the constraint equations as well as $2-c_{14}$ and $2c_1-c_+c_-$ for the wave equations. We have therefore implicitly excluded cases where these quantities are equal to zero. If we were to fine-tune these combinations so that they vanish then the derivation of certain modes falls apart. Since we demand that our solutions have no static or homogeneous terms, these modes are most often constrained to zero instead of being described by a wave equation.\footnote{
There is only one exception to this: if $c_+=0$, the constraint equation for $v_i$ falls apart and $\gamma_i$ is constrained to zero, but $v_i$ still obeys a wave equation with mode speed $c_V$. However, since we do not consider direct couplings between aether and matter in this paper, this kind of wave has no measurable effect.
} Similar issues arise in the cases $1-c_+=0$ as well as $2+3c_2+c_+=0$ and $2c_1-c_+c_-=0$. All three cases result in vanishing inverse mode speeds,
changing the character of the equation from hyperbolic to elliptic. Therefore, we now explicitly exclude these cases as well.

A possible solution to each wave equation is given by a monochromatic plane wave
that propagates along the $z$ direction:
\begin{equation}
\psi = \hat\psi \exp\left[i \omega \left(\frac{z}{c_{p}}-t\right)-\frac{i \pi}{2}\right]
\label{eq:def_psi}
\end{equation}
where the subscript $p$ indicates that the propagation speed corresponds to a scalar, vector, or tensorial polarization.
Here, $\hat\psi$ is the wave amplitude, $\omega$ is the angular frequency,
and $-i\pi/2$ is a phase
chosen such that the real part of Eq.~\eqref{eq:def_psi} is a sine of $\omega(z-t)$. This means that $\psi = 0$ for $z=t$ and, specifically, $z=t=0$. We choose this convention so that the coordinates of test particles in the $xy$ plane are unaffected at $z=t=0$.

Using
Eq.~\eqref{eq:def_psi}, we can construct the full metric and aether perturbations, which expressed in terms of the SVT decomposition read
\begin{align}
	&
	w^\mu=\left(
	\begin{array}{c}
		\dfrac{h_{00}}{2}\\\\v_i
	\end{array}
	\right),\;
	&
	h_{\mu\nu}=
	\left(
	\begin{array}{ccc}
		h_{00}&&\gamma_i\\\\
		\gamma_i&&\phi_{ij}+\dfrac{1}{2}P_{ij}f+\phi_{,ij}
	\end{array}
	\right).
    \nonumber \\
    \label{eq:wmu_and_hmunu}
\end{align}

\begin{figure}[t]
\centering
\includegraphics[width=0.75\linewidth]{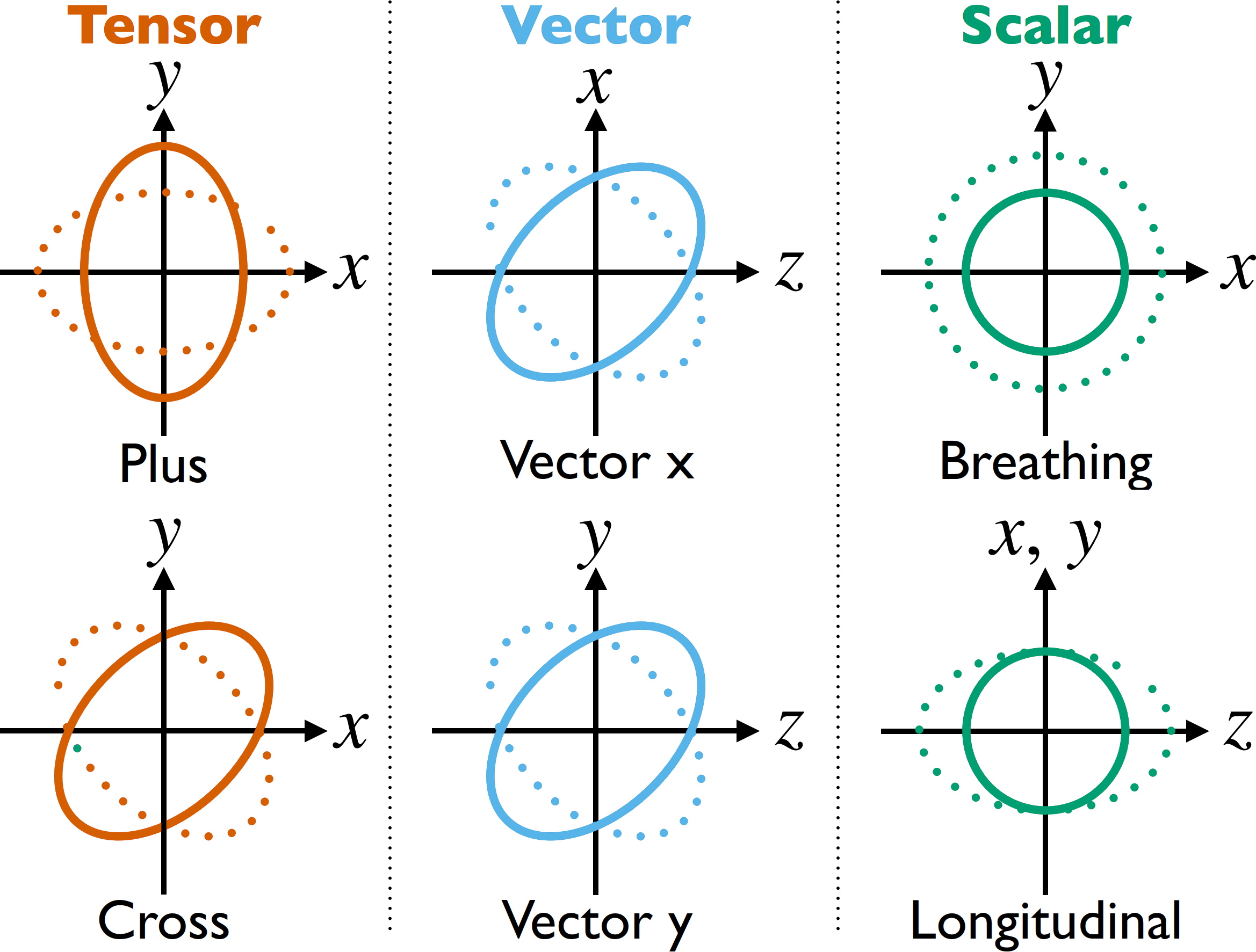}
\caption[Effect of Gravitational Waves beyond General Relativity]{Qualitative effect of GWs beyond GR on circles and spheres composed of test particles. We assume the GW wavelengths to be much larger than the diameter of any such circle or sphere.
Tensor and vector waves deform circles into ellipses in different planes and along different axes. These modes preserve the circle areas. The scalar modes instead expand and contract circles and spheres, changing their area and volume, respectively, in the $xy$ plane for the breathing mode and in the $z$ direction for the longitudinal mode. Figure is modified from Ref.~\cite{Takeda_2020}; compare also Ref.~\cite{Nishizawa:2009bf}.}
\label{fig:beyond_gr_modes}
\end{figure}

If we specify the $z$ direction as the propagation direction, then there are six gauge invariant polarizations for GWs in theories beyond GR:
\begin{enumerate}
\item the scalar breathing mode $h_b$,
\item the scalar longitudinal mode $h_l$,
\item the vector $x$ mode $h_x$,
\item the vector $y$ mode $h_y$,
\item the tensor plus mode $h_+$,
\item the tensor cross mode $h_\times$.
\end{enumerate}
The effect of all six polarizations on a ring of test particles is shown in Fig.~\ref{fig:beyond_gr_modes}.

\begin{figure*}[!th]
	\centering
	\includegraphics[width=0.95\linewidth]{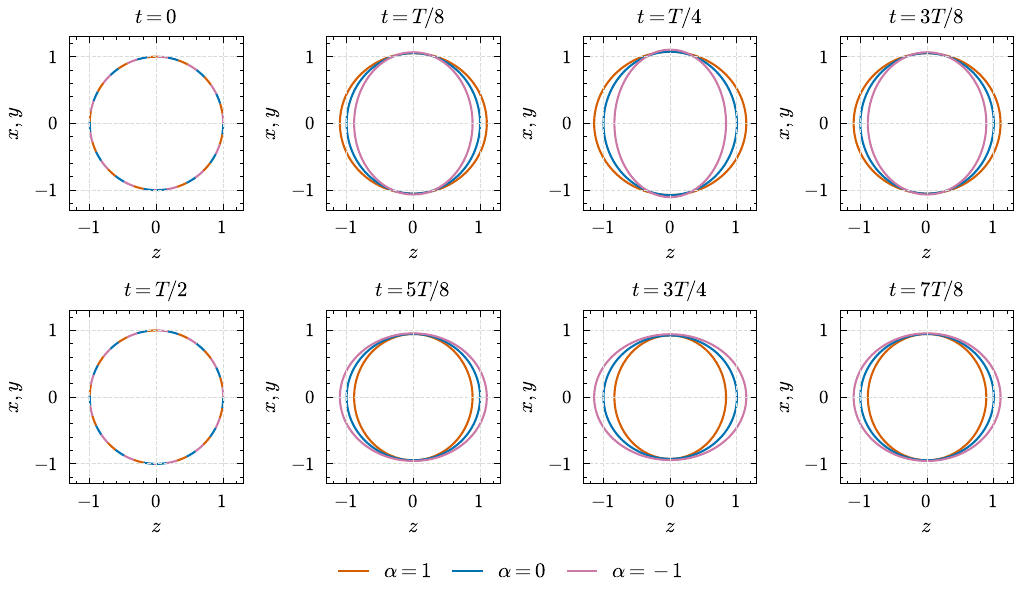}
	\caption[Effect of Scalar Waves in Einstein-Aether Theory]{Effect of scalar waves with amplitude 0.3 and different values of $\alpha$ on a sphere of test particles with diameter $1$ and a scalar wavelength much larger than $1$.
    Time is measured in fractions of the period $T$, and
    different curves correspond to $\alpha = 0$ and $\pm 1$. The sphere expands and contracts both in the $xy$ plane and in the $z$ direction. The effect differs based on the value of $\alpha$. For example, at $t=T/4$ the sphere expands in the $z$ direction for $\alpha=1$, while it contracts for $\alpha=-1$.}
	\label{fig:scalar_mode}
\end{figure*}

In Einstein-aether theory the vector modes are proportional to $\gamma_1$ and $\gamma_2$ and the tensor modes are proportional to $\phi_{11}$ and $\phi_{12}$. The breathing mode is proportional to $F$ and the longitudinal mode is proportional to both $F$ and a linear combination of theory constants that we call $\alpha$:
\begin{align}
\alpha= \frac{(1-c_+)(2+3c_2+c_+)}{c_{123}(2-c_{14})}-\frac{1+c_2}{c_{123}}.
\end{align}
This means that $h_b$ and $h_l$ are combined into a joint scalar polarization. Whether or not longitudinal effects are in phase with breathing effects, are out of phase, or vanish completely depends on the value of $\alpha$.
In Fig.~\ref{fig:scalar_mode} we show the effect of the joint breathing and longitudinal
scalar polarization on a sphere of test particles for $\alpha = 0$ and $\pm 1$.

A more detailed derivation of the effects of Einstein-aether waves can be found in Appendix~\ref{app:einstein_aether_waves}.

\subsection{Theory constraints}\label{subsec:theory_constraints}

Theory and observations impose a variety of constraints on Einstein-aether theory. The theory allows for scalar, vector and tensor propagating modes with speeds that differ from $1$. To avoid imaginary frequency instabilities, the squared speeds must be non-negative, $c_{S,V,T}^2\geq0$~\cite{Jacobson:2004ts,Jacobson:2007veq}. To avoid the existence of gravitational Cherenkov radiation, they can also not be smaller than $1$ up to a precision of about $10^{-15}$, $c_{S,V,T}^2> 1-10^{-15}$~\cite{Elliott:2005va}.\footnote{This is equivalent to $c_{S,V,T}>1-5\times10^{-16}$.} This constraint can equivalently be obtained by considerations on the weak cosmic censorship conjecture under the test particle approximation: speeds $c_T^2<1$ would allow for the overcharging of extremal charged Einstein-aether black holes~\cite{Ghosh:2021cub}.\footnote{In Ref.~\cite{Ghosh:2021cub} the constraint is stated to be $0\leq c_+ < 1$. This corresponds to $c_T^2 \geq 1$, without the window of width $10^{-15}$ below $1$ allowed for by the Cherenkov constraint. However, since said reference utilizes the test particle approximation, we will assume the result to be not perfectly precise so that the small window is still open.} In addition to the constraints on the mode speeds, the Einstein-aether mode energy densities must be positive~\cite{Eling:2005zq} and the coefficients of the time kinetic term must be positive so as to avoid ghosts~\cite{Oost:2018tcv}.

Multimessenger observations from the GW signal GW170817 and the gamma-ray burst GRB 170817A give us the best current observational constraint on the theory. They constrain the speed of tensor waves to be above $1-3\times 10^{-15}$ and below $1+7\times10^{-16}$~\cite{Gong:2018cgj,Oost:2018tcv}. However, $c_T$ depends only on $c_+$, which means that the “parameter plane” spanned by $c_1 + c_3 = 0$ remains unconstrained.
Part of the motivation for studying GW lensing in Einstein-aether theory is a hope of breaking this degeneracy and restricting parameters directly.

Further observational constraints can be imposed on the Newtonian gravitational constant $G_N$ and cosmological gravitational constant $G_{\text{cosmo}}$, which in Einstein-aether theory no longer have the same value~\cite{Carroll:2004ai,Jacobson:2007veq}. The simple observation that gravity is attractive demands $G_N>0$, although it can be shown that this constraint is automatically included in the
constraint from gravitational Cherenkov radiation~\cite{Jacobson:2007veq}. Observations of primordial $^4$He abundance give rise to a so-called nucleosynthesis constraint that demands $\abs{G_{\text{cosmo}}/G_N}-1\lesssim 1/8$~\cite{Carroll:2004ai,Jacobson:2007veq}.

Another group of observational constraints stems from parametrized post-Newtonian analysis~\cite{Nordtvedt:1968qs,Nordtvedt:1969zz,Will:1971zzb,Will:1972zz}.
The parametrized post-Newtonian formalism contains a set of parameters called preferred frame parameters, $\alpha_1$ and $\alpha_2$. They have been constrained  to $\abs{\alpha_1}\leq 10^{-4}$ and $\abs{\alpha_2}\leq 10^{-7}$ by using lunar laser ranging and the solar alignment with the ecliptic~\cite{Will:1993hxu,Muller:1996up,Graesser:2005bg}.
These parameters are zero in GR, but, in general, are nonvanishing in Einstein-aether theory. In the strong field regime there exist equivalent parameters $\hat\alpha_1$ and $\hat\alpha_2$ which can be constrained to $\abs{\hat\alpha_1}\leq 3.5\times 10^{-5}$ and $\abs{\hat\alpha_2}\leq 1.6\times 10^{-9}$ at 95\% confidence using observations of isolated millisecond pulsars~\cite{Shao:2012eg,Shao:2013wga,Oost:2018tcv}.

The final constraint is provided by observations of the orbital decay of binary and trinary pulsars. By combining them with post-Newtonian calculations, these observations can provide constraints for $c_+$ and $c_-$~\cite{Foster:2007gr,Yagi:2013ava}, but only under the assumption of small $\alpha_{1,2}$ compared to $c_+$ and $c_-$. Now that the best constraint provided by multimessenger observations is much stricter than the conditions on the preferred frame parameters these constraints have become obsolete~\cite{Oost:2018tcv,Sarbach:2019yso}. However, $\alpha_1$ can still be constrained with pulsar observations. The constraint obtained from trinary pulsars is about an order of magnitude stricter than that from lunar laser ranging: $\abs{\alpha_1}\leq 10^{-5}$~\cite{Gupta:2021vdj}.

\section{Gravitational wave lensing}\label{sec:gw_lensing}

We now study Einstein-aether theory with a non-Minkowski background. First, in Sec.~\ref{subsec:approx_field_eq}, we apply some approximations and obtain the new field equations. Solving the field equations is not trivial, hence, in Sec.~\ref{subsec:propagation_eigenstate_framework}, we apply to the propagation eigenstate framework~\cite{Ezquiaga:2020dao}. This framework allows us to study the mixing of modes resulting from GW propagation on general backgrounds. The central object of the framework is the so-called kinetic matrix. Eigenvalues and eigenvectors of the kinetic matrix will be crucial, so we perform an eigendecomposition in Sec.~\ref{subsec:kinetic_matrix_eigendecomposition}. Finally, in Sec.~\ref{subsec:propagation_eigenstate_speeds}, we apply the framework to obtain the speeds of the lensed and mixed propagating states.

\subsection{Approximations and field equations}\label{subsec:approx_field_eq}

After reviewing the standard case, i.\,e., GWs in Einstein-aether without lensing and with a flat background, we want to focus on the extension where GW lensing leads to a mixing between modes. To introduce lensing, instead of using an altogether different background, we leave the Minkowski background intact, but no longer assume that the background aether is $\delta^\mu_0$, instead allowing it to have spatial components. This is an approximation, the exact meaning of which is best understood through an expansion of the metric tensor in Riemann normal coordinates around a point $0$~\cite{Muller:1997zk}:
\begin{align}
g_{\alpha\beta}\left(x^\mu\right) &= g_{\alpha\beta}(0) + \frac{1}{3} R_{\alpha\mu\nu\beta}(0) x^\mu x^\nu + \mathcal{O}\left(x^3\right).
\end{align}

We find that the leading order correction to spacetime being flat is given by the curvature at $0$ and the distance $x^\mu$ we have traveled from $0$.
Roughly speaking, our approximation holds if curvature is small on the scale of lensing. Physically, this corresponds to passing far from a point lens rather than close by it, in the regime where its background field is weak. We focus on a perturbation of the background aether because, while perturbations of the metric due to lensing have been studied before (see, e.g., Refs.~\cite{Saltas:2014dha,Sawicki:2016klv,Amendola:2017orw}), perturbations of the aether field have not.
Since the correction to the aether will then also be small, we go only to first order in the spatial components of the aether.

We will call the new, nontrivial background aether $\mu^\alpha$. We again linearize around the contravariant background:
\begin{align}
\overline{u}^\alpha &= \mu^\alpha + w^\alpha.
\end{align}
The aether background represents a generic inhomogeneous configuration, sourced by the matter distribution, e.g.~a massive body such as a galaxy or a cosmological perturbation sourced by a large-scale structure. Linearization prevents only the study of compact objects, such as black holes and neutron stars, where the aether field could have order unity deviations~\cite{Eling:2007xh}. Even then, only signals propagating close to the gravitational radius will require going beyond our framework.

We will make one more approximation: that derivatives of $\mu^\alpha$ are small compared to derivatives of $w^\alpha$. Formally, this means we are restricting ourselves to the limit of geometric optics or ray optics. In this limit, the wavelengths of the GWs are small compared with the length scale of the lens set by its mass $M_{L}$. The frequency range of LIGO is between 10\,Hz and\,10 kHz~\cite{LIGO_2015}. Assuming that GWs propagate roughly with the speed of light, $\lambda_{\text{GW}}=c/f_{\text{GW}}$ with wavelength $\lambda_{\text{GW}}$ and frequency $f_{\text{GW}}$ gives us a wavelength range of between 30 and 30,000\,km, with typical wavelengths in the range of 1,000\,km. While black holes, planets or stars may be too small for our approximation to hold, galaxies serve as ideal lenses for this approximation. Even when viewed orthogonal to the disk, the Milky Way is about $10^{13}$ times bigger than the characteristic wavelength, $\sim$ 1,000\,km.\footnote{According to Bland-Hawthorn and Gerhard~\cite{Bland_Hawthorn_2016}, the exponential scale height of the dominant old thin disk of the Milky Way galaxy at the sun's location is between 220 and 450 parsecs, which corresponds to approximately $10^{19}$\,m.}

The validity of the high-frequency (eikonal) expansion also requires that the time delays associated with the lens are larger than the GW period,
that is,
\begin{equation}
f\gg \frac{1}{8\pi GM_L(1+z_L)} \sim 100 \text{Hz}\left(\frac{100M_\odot}{(1+z_L)M_L}\right)\,,
\end{equation}
where $(1+z_L)M_L$ is the redshifted effective mass of the lens~\cite[Sec.~IIA]{Tambalo:2022wlm}.
For extended lenses with size $\gg GM$ this is a more restrictive requirement. If it is not fulfilled, a wave-optic treatment is necessary to describe GW lensing~\cite{Schneider:1992,Takahashi:2003ix, Leung:2023lmq,Tambalo:2022plm,Villarrubia-Rojo:2024xcj}. Ground and space GW observatories satisfy the eikonal condition for galactic and cluster lenses $M_L\geq 10^{11}M_\odot$, which dominate the lensing probability~\cite{Brando:2024inp}. A wave-optic treatment of galaxy lenses is only necessary for nano Hertz GWs~\cite{Jow:2024bwq}.

With this in mind, we can already transfer some results from the unlensed case. Consider calculating a linearized covariant derivative of the aether, $\overline{D_\gamma u^\rho}$. The derivative operator $D_\gamma$ has zeroth and first order contributions: partial derivatives and linearized Christoffel symbols, respectively. Similarly, the aether has its background configuration $\mu^{\rho}$ and
a perturbation $w^{\rho}$ to it. The combination of both first order terms would be of second order, leaving us with three nonvanishing terms:
\begin{align}
\overline{D_\gamma u^\rho}
&=
\partial_\gamma \mu^\rho
+ \partial_\gamma w^\rho
+ \overline\Gamma^{\rho}_{\gamma\sigma} \mu^\sigma.
\label{eq:covariant_aether_derivative}
\end{align}

Since we assume $\partial_\gamma \mu^\rho\ll \partial_\gamma w^\rho$ the first term drops out, leaving us with only contributions of first order in the perturbation. This means that just as in the unlensed case products of aether, aether acceleration, and aether current must drop out and any term multiplied with them must be of zeroth order:
\begin{align}
\overline{T}^{\alpha\beta}_{\text{ae}}&=
\mu^\mu \partial_\mu \overline{J}^{(\alpha\beta)}
- \mu^{(\beta} \partial_\mu \overline{J}^{\alpha)\mu},\\
\overline{\text{\AE}^{[0} u^{i]}}&= \mu^{[i} \partial_\mu \overline{J}^{\mu|0]},\label{eq:bonus_tensor_with_lensing}
\end{align}
that, by inserting the aether current, can be rewritten as
\begin{align}
\overline{T}^{\alpha\beta}_{\text{ae}}&=
M^{(\alpha|\gamma(0)}_{\phantom{(\alpha|\gamma}\nu\rho}
\left[\eta^{\beta)\nu} \mu^\mu-\eta^{\mu\nu} \mu^{\beta)} \right]
\partial_\mu \overline{D_\gamma u^\rho},\label{eq:lensed_linearized_aether_stress_energy_tensor}
\\
\overline{\text{\AE}^{[0} u^{i]}}
&=
M^{\mu\gamma(0)}_{\phantom{\mu\gamma}\nu\rho}  \mu^{[i} \eta^{0]\nu} \partial_\mu \overline{D_\gamma u^\rho}.\label{eq:lensed_linearized_bonus_tensor}
\end{align}

To summarize, we make the following three physical approximations:
\begin{enumerate}
	\item linearized GW approximation: the GW amplitudes are small,
	\item weak background field approximation: the GWs pass far from the lens,
	\item geometric optics (eikonal) approximation: the lens is much larger than the GW wavelength.
\end{enumerate}

Mathematically, these three approximations mean that
\begin{enumerate}
	\item we linearize in $h_{\mu\nu}$ and $w_\mu$,
	\item we keep the Minkowski background and linearize in $\mu_i$,
	\item we neglect $\partial_\gamma \mu^\rho$ compared to $\partial_\gamma w^\rho$,
\end{enumerate}
respectively.

Once again our first step, after linearization, is to eliminate $w^0$ from the field equations by using the constraint equation. The linearized constraint equation takes the following form:
\begin{align}
0
&=
\overline{g_{\alpha\beta} u^\alpha u^\beta} + 1
\nn\\&=
(\mu_k)^2
-(\mu_0)^2
+h_{00} (\mu_0)^2
+2\gamma_k \mu_k \mu^0
\nn\\&\mathrel{\phantom{=}}
+\, \phi_{kl} \mu_k \mu_l
+\frac{1}{2} (\mu_k)^2 F
-\frac{1}{2} U^2 f
+ U^2 \phi
\nn\\&\mathrel{\phantom{=}}
- \, 2 w^0 \mu^0
+ 2 v_k \mu_k
+1.
\label{eq:constraint_tmp}
\end{align}
Here we defined $U \psi= \mu_k \psi_{,k}$, for an arbitrary scalar function $\psi$.
We now go to first order in $\mu_k$, which means $(\mu_0)^2=1$ and $(\mu_k)^2=0$. Equation~\eqref{eq:constraint_tmp} then simplifies to
\begin{align}
0&=h_{00}+2\gamma_k \mu_k -2 w^0+ 2 v_k \mu_k,
\end{align}
which we can solve for $w^0$:
\begin{align}
	w^0
	&=
	\frac{1}{2} h_{00}
	+\mu_k \gamma_k
	+\mu_k v_k.
\end{align}
In the trivial case $\mu_k=0$ and we reclaim
the trivial constraint equation $w^0=h_{00}/2$; cf.~Eq.~\eqref{eq:wmu_and_hmunu}.

We now proceed as we did previously: we calculate the linearized aether stress-energy tensor [Eq.~\eqref{eq:lensed_linearized_aether_stress_energy_tensor}], $\overline{\text{\AE}^{[0} u^{i]}}$ [Eq.~\eqref{eq:lensed_linearized_bonus_tensor}], and, finally, the rather lengthy components of the effective tensor [Eqs.~\eqref{eq:X00} through~\eqref{eq:Xij}]. We find
\begingroup
\allowdisplaybreaks
\begin{align}
X_{00}
&=
 \frac{1}{2} c_{14} \Delta h_{00}
-\frac{1}{2} \Delta F
+\frac{1}{2} c_2 U \dot{F}
\nn\\&\mathrel{\phantom{=}}
-\frac{1}{2} (c_--c_2+2c_4) U \Delta \dot\phi
\nn\\&\mathrel{\phantom{=}}
+\frac{1}{2} (c_++2c_4) \mu_k \Delta \gamma_k
+c_3 \mu_k \Delta v_k,
\\[5mm]
X_{0i}
&=
 \frac{1}{2} c_{14} \dot{h}_{00,i}
-\frac{1}{4} \left(c_++2c_4\right) \mu_i \Delta h_{00}
\nn\\&\mathrel{\phantom{=}}
+\frac{1}{4} (c_++2c_4) U h_{00,i}
-\frac{1}{2} \dot{F}_{,i}
+\frac{1}{4} c_- \mu_i \Delta F
\nn\\&\mathrel{\phantom{=}}
-\frac{1}{4} (c_--2c_2+2c_4) \mu_i \ddot{F}
-\frac{1}{4} c_- U F_{,i}
\nn\\&\mathrel{\phantom{=}}
+\frac{1}{4} (c_-+2c_4) U \ddot{f}_{,i}
+\frac{1}{2} c_2 \mu_i \Delta \ddot\phi
\nn\\&\mathrel{\phantom{=}}
-\frac{1}{2} (c_-+2c_4) U \ddot\phi_{,i}
-\frac{1}{2} (1-c_-) \Delta \gamma_i
-c_{14} \ddot\gamma_i
\nn\\&\mathrel{\phantom{=}}
+\frac{1}{2} (c_++2c_4) \mu_k \dot\gamma_{k,i}
-(c_++2c_4) U \dot\gamma_i
\nn\\&\mathrel{\phantom{=}}
+\frac{1}{2} c_- \Delta v_i
-c_{14} \ddot{v}_i
+c_3 \mu_k \dot{v}_{k,i}
\nn\\&\mathrel{\phantom{=}}
-\frac{1}{2} (c_++4c_4) U \dot{v}_i
+\frac{1}{2} c_- \mu_k \Delta \phi_{ik}
\nn\\&\mathrel{\phantom{=}}
-\frac{1}{2} (c_-+2c_4) \mu_k \ddot\phi_{ik},
\\[5mm]
X_{i0}
&=
-\frac{1}{4} \left(c_-+2c_4\right) \mu_i \Delta h_{00}
\nn\\&\mathrel{\phantom{=}}
+\frac{1}{4} (c_++2c_2) U h_{00,i}
-\frac{1}{2} (1+c_2) \dot{F}_{,i}
\nn\\&\mathrel{\phantom{=}}
-\frac{1}{2} c_2 U F_{,i}
-\frac{1}{2} c_{123} \Delta \dot\phi_{,i}
-\frac{1}{2} c_{123} U \Delta \phi_{,i}
\nn\\&\mathrel{\phantom{=}}
-\frac{1}{2} \Delta \gamma_i
-c_2 \mu_k \dot\gamma_{k,i}
-\frac{1}{2} c_+ U \dot\gamma_i
-\frac{1}{2} c_+ \Delta v_i
\nn\\&\mathrel{\phantom{=}}
-c_2 \mu_k \dot{v}_{k,i}
-\frac{1}{2} c_+ U \dot{v}_i,
\\[5mm]
X_{ij}
&=
-\frac{1}{2} P_{ij} h_{00}
+\frac{1}{4} (c_3-c_4) \mu_i \dot{h}_{00,j}
\nn\\&\mathrel{\phantom{=}}
+\frac{1}{4} (c_3-c_4) \mu_j \dot{h}_{00,i}
+\frac{1}{2} c_2 \delta_{ij} U \dot{h}_{00}
\nn\\&\mathrel{\phantom{=}}
+\frac{1}{4} P_{ij} F
-\frac{1}{4} (c_++2c_2) \delta_{ij} \ddot{F}
-\frac{1}{4} P_{ij} \ddot{f}
\nn\\&\mathrel{\phantom{=}}
+\frac{1}{4} (c_+-2) \ddot{f}_{,ij}
+\frac{1}{4} c_2 \mu_i \dot{F}_{,j}
+\frac{1}{4} c_2 \mu_j \dot{F}_{,i}
\nn\\&\mathrel{\phantom{=}}
-\frac{1}{2} (c_++2c_2) \delta_{ij} U \dot{F}
+\frac{1}{2} c_+ U \dot{f}_{,ij}
-\frac{1}{2} P_{ij} \ddot\phi
\nn\\&\mathrel{\phantom{=}}
-\frac{1}{2} c_2 \delta_{ij} \Delta \ddot\phi
-\frac{1}{2}c_+ \ddot\phi_{,ij}
+\frac{1}{4} c_{123} \mu_i \Delta \dot\phi_{,j}
\nn\\&\mathrel{\phantom{=}}
+\frac{1}{4} c_{123} \mu_j \Delta \dot\phi_{,i}
-c_2 U \delta_{ij} \Delta \dot\phi
-c_+ U \dot\phi_{,ij}
\nn\\&\mathrel{\phantom{=}}
- \frac{1}{2} \dot{\gamma}_{i,j}
- \frac{1}{2} \dot{\gamma}_{j,i}
-\frac{1}{4} c_- \mu_i \Delta \gamma_j
-\frac{1}{4} c_- \mu_j \Delta \gamma_i
\nn\\&\mathrel{\phantom{=}}
-\frac{1}{2} (c_3-c_4) \mu_i \ddot\gamma_j
-\frac{1}{2} (c_3-c_4) \mu_j \ddot\gamma_i
\nn\\&\mathrel{\phantom{=}}
-c_2 \delta_{ij} \mu_k \ddot\gamma_k
-\frac{1}{2} c_+ \dot{v}_{i,j}
-\frac{1}{2} c_+ \dot{v}_{j,i}
\nn\\&\mathrel{\phantom{=}}
+\frac{1}{2} c_3 \mu_i \Delta v_j
+\frac{1}{2} c_3 \mu_j \Delta v_i
-\frac{1}{2} (c_3-c_4) \mu_i \ddot{v}_j
\nn\\&\mathrel{\phantom{=}}
-\frac{1}{2} (c_3-c_4) \mu_j \ddot{v}_i
-c_2 \delta_{ij} \mu_k \ddot{v}_k
-\frac{1}{2} c_+ U v_{i,j}
\nn\\&\mathrel{\phantom{=}}
-\frac{1}{2} c_+ U v_{j,i}
- \frac{1}{2} \Delta\phi_{ij}
+ \frac{1}{2} (1-c_+) \ddot\phi_{ij}
\nn\\&\mathrel{\phantom{=}}
-c_+ U \dot\phi_{ij},
\end{align}
\endgroup

The Helmholtz decomposition of $X_{0i}$ and $X_{i0}$ requires the decomposition of $\mu_i$ into transverse and longitudinal parts, but is otherwise simple. The SVT decomposition of $X_{ij}$ is less straightforward: some terms simply do not fit into our scheme of having every term be transverse traceless, be a symmetrized derivative of a transverse vector, or be expressed in terms of $\partial_i \partial_j$ or $P_{ij}$. However, while the SVT decomposition is difficult if not impossible in real space, it becomes possible in Fourier space.\footnote{Specifically, the Fourier space SVT decomposition contains terms proportional to $k^{-2}$. These terms are the problem with a real space SVT decomposition: they are nonlocal and can only be described with inverse operators.} We
work in Fourier space to proceed. We define the Fourier transform $\mathcal{F}$ and the inverse Fourier transform $\mathcal{F}^{-1}$ of a function $\psi$ as
\begin{align}
\mathcal{F}[\psi] &= \int {\rm d}^4 x \, \psi e^{-i x_\mu k^\mu},
\\
\mathcal{F}^{-1}[\psi] &= \frac{1}{2\pi} \int {\rm d}^4k \, \psi e^{i x_\mu k^\mu},
\end{align}
and obtain the same four constraints we utilized previously up to first order in $\mu_k$:
\begin{align}
h_{00}
&=
\frac{F}{c_{14}}
+\frac{2c_{14}(1+c_2)-c_{123}}{c_{14}c_{123}} \frac{\omega \mu_l k_l}{k^2} F
\nn\\&\mathrel{\phantom{=}}
+ \frac{2c_3-(c_++2c_4) c_+}{c_{14} c_+} \mu_k \gamma_k
\label{eq:linear_const_h00}
\\
\phi
&=
\frac{1+c_2}{c_{123}} \frac{F}{k^2}
+ \frac{1}{c_{14}} \frac{\mu_l k_l}{\omega k^2} F
- \frac{2c_2 (1-c_+)}{c_{123} c_+} \mu_k \gamma_k,
\\
v_i
&=
-\frac{c_-+2c_4}{2 c_+ c_{14}} \mu_i^T F
-\frac{\gamma_i}{c_+}
-\frac{1-c_+}{c_+} \frac{\omega \mu_l k_l}{k^2} \gamma_i.
\label{eq:linear_const_vi}
\end{align}

The remaining equations will serve as the gateway to utilizing what is known as the propagation eigenstate framework~\cite{Ezquiaga:2020dao}
and finding the mixing between the modes. The transverse part of $X_{0i}=0$ carries the first two DOF:
\begingroup
\allowdisplaybreaks
\begin{align}
	0
&=
-\left(\frac{c_++2c_4}{2c_{14}}-\frac{c_-}{2} +c_- \frac{c_-+2c_4}{2 c_+ c_{14}}\right) \mu_i^T  k^2 F
\nn\\&\mathrel{\phantom{=}}
-\left(\frac{c_--2c_2+2c_4}{2}+c_2 \frac{1+c_2}{c_{123}}-\frac{c_-+2c_4}{c_+}\right) \mu_i^T \omega^2 F
\nn\\&\mathrel{\phantom{=}}
-\frac{2c_1-c_+c_-}{c_+} k^2\gamma_i
+\frac{2c_{14}(1-c_+)}{c_+} \left(1+\frac{\omega \mu_l k_l}{k^2}\right) \omega^2 \gamma_i
\nn\\&\mathrel{\phantom{=}}
+\left[2(c_++2c_4)-\frac{c_++4c_4}{c_+}-c_- \frac{1-c_+}{c_+}\right] \omega \mu_l k_l \gamma_i
\nn\\&\mathrel{\phantom{=}}
+c_- \mu_k k^2 \phi_{ik}
-(c_-+2c_4) \omega^2 \mu_k \phi_{ik}.
\end{align}
\endgroup
Inserting the constraints into the transverse traceless part of $X_{ij}=0$ yields equations for two more DOF:
\begingroup
\allowdisplaybreaks
\begin{align}
0
&=
-\frac{2c_3+c_+c_-}{2c_+} k^2 \left(\mu_i^T \gamma_j+\mu_j^T \gamma_i\right)
\nn\\&\mathrel{\phantom{=}}
+\frac{2c_3+c_+c_-}{2c_+} \left(\delta_{ij} k^2 -k_i k_j \right) \mu_k  \gamma_k
\nn\\&\mathrel{\phantom{=}}
+ \frac{(1-c_+)(c_3-c_4)}{c_+} \omega^2 \left(\mu_i^T \gamma_j+\mu_j^T \gamma_i\right)
\nn\\&\mathrel{\phantom{=}}
- \frac{(1-c_+)(c_3-c_4)}{c_+} \frac{\omega^2}{k^2} \left(\delta_{ij} k^2 -k_i k_j \right) \mu_k \gamma_k
\nn\\&\mathrel{\phantom{=}}
-k^2\phi_{ij}
+(1-c_+)\omega^2 \phi_{ij}
+2c_+ \omega \mu_l k_l \phi_{ij}.
\end{align}
Finally, we can insert our constraints into the transverse trace part of $X_{ij}=0$:
\begin{align}
0
&=
\left(\frac{1}{c_{14}}-\frac{1}{2}\right) F
-\frac{(1-c_+)(2+c_++3c_2)}{2c_{123}} \frac{\omega^2}{k^2} F
\nn\\&\mathrel{\phantom{=}}
-\left[
c_++2c_2
-\frac{2(1+c_2)^2}{c_{123}}
+ \frac{2}{c_{14}}
\right]
\frac{\omega \mu_l k_l}{k^2} F
\nn\\&\mathrel{\phantom{=}}
+\left[\frac{2c_3-(c_++2c_4) c_+}{c_{14} c_+}
+\frac{c_-}{2}
+ \frac{c_3}{c_+} \right] \mu_k \gamma_k
\nn\\&\mathrel{\phantom{=}}
+
\bigg[
\frac{2c_2 (1-c_+) (1+c_2)}{c_{123} c_+}
\nn\\&\mathrel{\phantom{=+\bigg[}}
+(2 c_2+c_3-c_4) \left(1-\frac{1}{c_+}\right)
\bigg]
\frac{\omega^2}{k^2} \mu_k \gamma_k
.
\end{align}
\endgroup

\subsection{Propagation eigenstate framework}\label{subsec:propagation_eigenstate_framework}

To solve our field equations with the propagation eigenstate framework, we must first

\begin{enumerate}
\item normalize our equations so the trivial limit is of the form $0=c_p^2\Delta \psi-\ddot\psi$;
\item convert our equations into a matrix equation.
\end{enumerate}

We now explicitly consider the $z$ direction as the propagation direction. In this case, all transverse vectors have components only in the $x$ or $y$ direction. Applying this process, we arrive at a matrix equation containing a 5$\times$5 matrix:

\begin{widetext}
\begin{align}
\left(\begin{array}{ccccccccc}
K_{SS}&&K_{SV} \mu_1&&K_{SV} \mu_2&&0&&0\\\\
K_{VS} \mu_1&&K_{VV}&&0&&K_{VT} \mu_1 &&K_{VT} \mu_2\\\\
K_{VS} \mu_2&&0&&K_{VV}&&-K_{VT} \mu_2 &&K_{VT} \mu_1\\\\
0&&K_{TV} \mu_1&&-K_{TV}\mu_2&&K_{TT}&&0\\\\
0&&K_{TV}\mu_2&&K_{TV} \mu_1&&0&&K_{TT}
\end{array}\right)
\left(\begin{array}{c}
F\\\\\gamma_1\\\\\gamma_2\\\\\phi_{11}\\\\\phi_{12}
\end{array}
\right)&=0.\label{eq:kinetic_matrix}
\end{align}
\end{widetext}

The entries of the matrix are second-order derivative operators:
\begingroup
\allowdisplaybreaks
\begin{align}
K_{SS}&=\omega^2 - c_S^2 k^2 + d_1 \omega k \mu_3,\label{eq:KSS}\\
K_{SV}&=-d_2 k^2-d_3 \omega^2,\\
K_{VS}&=-a_1 k^2-a_2 \omega^2,\\
K_{VV}&=\omega^2-c_V^2 k^2 + \frac{\omega^3}{k} \mu_3+a_3 \omega k \mu_3,\\
K_{VT}&=-a_4 k^2 -a_5 \omega^2,\\
K_{TV}&=-b_1 k^2-b_2 \omega^2,\label{eq:KTV}\\
K_{TT}&=\omega^2-c_T^2 k^2 +b_3 \omega k \mu_3.
\end{align}
\endgroup
Each coefficient is given by the theory parameters as follows:
\begingroup
\allowdisplaybreaks
\begin{align}
a_1&=\frac{c_+}{2c_{14} (1-c_+)}\bigg(\frac{c_++2c_4}{2c_{14}}-\frac{c_-}{2} +c_- \frac{c_-+2c_4}{2 c_+ c_{14}}\bigg),
\\
a_2&=\frac{c_+}{2c_{14} (1-c_+)}
\nn\\&\mathrel{\phantom{=}}\times
\left(\frac{c_--2c_2+2c_4}{2}+c_2 \frac{1+c_2}{c_{123}}-\frac{c_-+2c_4}{c_+}\right)
\\
a_3&=\frac{c_+}{2c_{14} (1-c_+)}
\nn\\&\mathrel{\phantom{=}}\times
\left[2(c_++2c_4)-\frac{c_++4c_4}{c_+}-c_- \frac{1-c_+}{c_+}\right],
\\
a_4&=-c_-\frac{c_+}{2c_{14} (1-c_+)},
\\
a_5&=(c_-+2c_4)\frac{c_+}{2c_{14} (1-c_+)},\\
b_1 &= \frac{2c_3+c_+c_-}{2c_+(1-c_+)},\\
b_2 &= - \frac{c_3-c_4}{c_+},\\
b_3 &= \frac{2c_+}{1-c_+},\\
d_1
&=
\frac{2c_{123}}{(1-c_+)(2+c_++3c_2)}
\nn\\&\mathrel{\phantom{=}}\times
\left[
c_++2c_2
-\frac{2(1+c_2)^2}{c_{123}}
+ \frac{2}{c_{14}}
\right],
\\
d_2
&=
\frac{2c_{123}}{(1-c_+)(2+c_++3c_2)}
\nn\\&\mathrel{\phantom{=}}\times
\left[\frac{2c_3-c_+(c_++2c_4)}{c_{14} c_+}
+\frac{c_-}{2}
+ \frac{c_3}{c_+} \right],
\\
d_3
&=
\frac{2c_{123}}{(1-c_+)(2+c_++3c_2)}
\nn\\&\mathrel{\phantom{=}}\times
\bigg[
\frac{2c_2 (1-c_+) (1+c_2)}{c_{123} c_+}
\nn\\&\mathrel{\phantom{=\times\bigg[}}
+(2 c_2+c_3-c_4) \left(1-\frac{1}{c_+}\right)
\bigg].\label{eq:d3}
\end{align}
\endgroup

For this matrix equation and related expressions we switch from index notation for tensors to using a more classical vector arrow and circumflex for vectors and matrices, respectively. We refer to the matrix as the kinetic matrix $\hat{K}$. The vector $(F,\gamma_1,\gamma_2,\phi_{11},\phi_{12})$ is called $\vec{h}$. Its five entries are called the interaction eigenstates.

Note that, if we were to go beyond leading order in the geometric-optics approximation, other matrices would appear in the total matrix equation: the amplitude matrix $\hat{A}$ containing first-order derivative operators and the mass matrix $\hat{M}$ containing no derivatives. In that case, the considerations that we apply to the kinetic matrix need to be applied to the sum of all three matrices $\hat{D}$~\cite{Dalang:2020eaj}.

Solving Eq.~(\ref{eq:kinetic_matrix}) is equivalent to finding combinations of fields that evolve independently and obtaining their dispersion relations, i.e., their speeds. The situation is analogous to neutrino flavor oscillations~\cite{Bilenky:1987ty}, where the field combinations that have well-defined propagation (i.e.~mass) do not have well-defined interactions, and vice versa. These field combinations are known as propagation and interaction eigenstates, respectively.

With $\hat{K}$ and $\vec{h}$ as a starting point we can now formally introduce the propagation eigenstate framework. Since $\hat{K}$ is not diagonal, the interaction eigenstates no longer obey simple wave equations.%
\footnote{The main difference with neutrinos is that their flavors mix through interactions with zero derivatives (i.e.,~mass matrix), rather than in their kinetic terms (which are diagonal).}
We now want to find the propagation eigenstates $\vec{H}$ that obey decoupled wave equations, the solutions of which describe propagation. To that end we seek to diagonalize $\hat{K}$. We do this by applying a matrix $\hat{M}$ to it so that $\hat{A} = \hat{M} \hat{K} \hat{M}^{-1}$ is diagonal. We can now modify the equation $\hat{K}\vec{h}=0$. We begin by multiplying with $\hat{M}$ from the left:
\begin{align}
\hat{M}\hat{K}\vec{h}=0.
\end{align}
Now, $\hat{M}^{-1}\hat{M}$ is the 5×5 identity matrix, which we can freely insert after $\hat{K}$:
\begin{align}
\hat{M}\hat{K}\hat{M}^{-1}\hat{M}\vec{h}=0.
\end{align}
We now define $\vec{H}=\hat{M}\vec{h}$:
\begin{align}
\hat{A}\vec{H}=0.
\end{align}
Therefore, we find that the propagation eigenstates are given by $\hat{M}\vec{h}$. The matrix $\hat{M}$ is called the mixing matrix because it describes how the interaction eigenstates mix into the propagation eigenstates. Since $\hat{M}$ diagonalizes $\hat{K}$ it contains the five eigenvectors of $\hat{K}$ as row vectors. The $n$th propagation eigenstate can then be calculated by taking the scalar product of the $n$th eigenvector of $\hat{K}$ with $\hat{h}$.

We also want to calculate with what speed the propagation eigenstates propagate. To that end we consider the entries of $\hat{A}$. They are the eigenvalues of $\hat{K}$. Since they describe wave equations it must be possible to bring them into the form $\omega^2-c_X^2 k^2$ with some mode speed $c_X$. By setting this to zero and solving for $\omega/k$ we can recover
$c_X$. In other words, setting the $n$th eigenvalue to zero and solving for $\omega/k$ yields the speed of the $n$th propagation eigenstate.

\subsection{Eigendecomposition of the kinetic matrix}\label{subsec:kinetic_matrix_eigendecomposition}

There are two equations that govern the eigendecomposition of a matrix.
The first one is the equation defining a pair of eigenvalue $E$ and eigenvectors $\vec{V}$:
\begin{align}
E\vec{V}&=\hat{K}\vec{V}.
\end{align}
The second one is the condition that the eigenvectors
are normalized. This means that the scalar product of an eigenvector with itself must be $1$:
\begin{align}
\vec{V}\cdot\vec{V}&=1.
\end{align}

We now split eigenvector, eigenvalue, and kinetic matrix into zeroth and first order in $\mu_k$. We will denote the zeroth order in $\mu_k$ by superscript $0\mu$, the first order by $1\mu$, and so on. The two equations we must solve can then also be split by order:
\begin{align}
E^{0\mu}\vec{V}^{0\mu}&=\hat{K}^{0\mu}\vec{V}^{0\mu},\\
E^{0\mu}\vec{V}^{1\mu}+E^{1\mu}\vec{V}^{0\mu}&=\hat{K}^{0\mu}\vec{V}^{1\mu}+\hat{K}^{1\mu}\vec{V}^{0\mu},\\
\vec{V}^{0\mu}\cdot\vec{V}^{0\mu}&=1,\\
\vec{V}^{0\mu}\cdot\vec{V}^{1\mu}&=0.
\end{align}

We note that $\hat{K}^{0\mu}$ only has diagonal elements. It follows from the zeroth-order equations that the zeroth-order eigenvalues
$E^{0\mu}$ are its
diagonal elements: These are $\omega^2-c_S^2 k^2$, two instances of $\omega^2-c_V^2 k^2$, and two instances of $\omega^2-c_T^2 k^2$. The corresponding zeroth-order eigenvectors are $\left(1,0,0,0,0\right)$, $\left(0,1,0,0,0\right)$, and so forth. As expected, without the background aether for mixing we recover the interaction eigenstates $F$, $\gamma_1$, $\gamma_2$, $\phi_{11}$, and $\phi_{12}$ with their respective speeds; cf.~Eqs.~\eqref{eq:cS}, \eqref{eq:cV}, and \eqref{eq:cT}.
We are left with the first-order equations which, as an example, we now solve for $\vec{V}^{0\mu}=\left(0,1,0,0,0\right)$. We first insert the zeroth-order eigenvalue and eigenvector into the first-order equations:

\begin{widetext}

\begingroup
\allowdisplaybreaks
\begin{align}
K_{VV}^{0\mu}
\left(\begin{array}{c}
V_1\\\\V_2\\\\V_3\\\\V_4\\\\V_5
\end{array}\right)^{1\mu}
+
E^{1\mu}
\left(\begin{array}{c}
	0\\\\1\\\\0\\\\0\\\\0
\end{array}\right)
\nn&=
\left(\begin{array}{ccccccccc}
K_{SS}&&0&&0&&0&&0\\\\
0&&K_{VV}&&0&&0&&0\\\\
0&&0&&K_{VV}&&0&&0\\\\
0&&0&&0&&K_{TT}&&0\\\\
0&&0&&0&&0&&K_{TT}
\end{array}\right)^{0\mu}
\left(\begin{array}{c}
	V_1\\\\V_2\\\\V_3\\\\V_4\\\\V_5
\end{array}\right)^{1\mu}
\nn\\[5mm]&\mathrel{\phantom{=}}+
\left(\begin{array}{ccccccccc}
	K_{SS}^{1\mu}&K_{SV} \mu_1&K_{SV} \mu_2&0&0\\\\
	K_{VS} \mu_1&K_{VV}^{1\mu}&0&K_{VT} \mu_1 &K_{VT} \mu_2\\\\
	K_{VS} \mu_2&0&K_{VV}^{1\mu}&-K_{VT} \mu_2 &K_{VT} \mu_1\\\\
	0&K_{TV} \mu_1&-K_{TV}\mu_2&K_{TT}^{1\mu}&0\\\\
	0&K_{TV}\mu_2&K_{TV} \mu_1&0&K_{TT}^{1\mu}
\end{array}\right)
\left(\begin{array}{c}
	0\\\\1\\\\0\\\\0\\\\0
\end{array}\right),
\label{eq:first_order_eigenequation}
\end{align}
\endgroup

\begin{align}
\left(\begin{array}{c}
	0\\\\1\\\\0\\\\0\\\\0
\end{array}\right)
\cdot
\left(\begin{array}{c}
	V_1\\\\V_2\\\\V_3\\\\V_4\\\\V_5
\end{array}\right)^{1\mu}
&=
0.
\end{align}

The second equation simply tells us that $V_2^{1\mu}=0$
, which we then insert
into
Eq.~\eqref{eq:first_order_eigenequation}
and simplify. These steps lead us to
\begin{align}
\left(\begin{array}{c}
K_{VV}^{0\mu}V_1^{1\mu}\\\\0\\\\K_{VV}^{0\mu}V_3^{1\mu}\\\\K_{VV}^{0\mu}V_4^{1\mu}\\\\K_{VV}^{0\mu}V_5^{1\mu}
\end{array}\right)
+
\left(\begin{array}{c}
	0\\\\E^{1\mu}\\\\0\\\\0\\\\0
\end{array}\right)
&=
\left(\begin{array}{c}
	K_{SS}^{0\mu} V_1^{1\mu}\\\\0\\\\K_{VV}^{0\mu} V_3^{1\mu}\\\\K_{TT}^{0\mu} V_4^{1\mu}\\\\K_{TT}^{0\mu} V_5^{1\mu}
\end{array}\right)
+
\left(\begin{array}{c}
	K_{SV}\mu_1\\\\K_{VV}^{1\mu}\\\\0\\\\K_{TV}\mu_1\\\\K_{TV}\mu_2
\end{array}\right).
\end{align}

\end{widetext}

The second row of the foregoing equation tells us that $E^{1\mu}=K_{TT}^{1\mu}$. The third row is trivially true, meaning that we have no information about $V_2^{1\mu}$. The remaining rows can be solved for $V_1^{1\mu}$, $V_4^{1\mu}$, and $V_5^{1\mu}$:
\begingroup\allowdisplaybreaks
\begin{align}
V_1^{1\mu}&=\frac{K_{SV}}{K_{VV}^{0\mu}-K_{SS}^{0\mu}}\mu_1,\label{eq:V1}\\
V_4^{1\mu}&=\frac{K_{TV}}{K_{VV}^{0\mu}-K_{TT}^{0\mu}}\mu_1,\\
V_5^{1\mu}&=\frac{K_{TV}}{K_{VV}^{0\mu}-K_{TT}^{0\mu}}\mu_2.\label{eq:V5}
\end{align}
\endgroup
These expressions are defined in terms of the elements of the kinetic matrix $\hat{K}$, which are defined in terms of $a_i$, $b_i$, and $d_i$ which in turn are defined in terms of the $c_i$; recall~Eqs.~\eqref{eq:KSS} through~\eqref{eq:d3}.
We can simplify Eqs.~\eqref{eq:V1} through~\eqref{eq:V5}
by inserting all these definitions into each other. Additionally, all fractions are multiplied with terms of first order in $\mu_k$. This means we can approximate the fractions to zeroth order in $\mu_k$ and replace $\omega^2/k^2$ with the correct mode speed. Since we are discussing a vector mode, we replace it with $c_V^2$. Carrying out these steps
we arrive at simple results, e.\,g.,~$V_4^{1\mu}=-\mu_1$, and we can calculate all five eigenvalues and eigenvectors:

\begin{widetext}
\begin{align}
E \in \bigg\{&
\omega^2-c_S^2 k^2+d_1 \omega k \mu_3,
\omega^2-c_V^2 k^2+\frac{\omega^3}{k} \mu_3 + a_3 \omega k \mu_3,
\omega^2-c_V^2 k^2+\frac{\omega^3}{k} \mu_3 + a_3 \omega k \mu_3,\nn\\&
\omega^2-c_T^2 k^2+b_3 \omega k \mu_3,
\omega^2-c_T^2 k^2+b_3 \omega k \mu_3
\bigg\},
\\[5mm]
\vec{V} \in&\left\{
\left(\begin{array}{c}1\\\\\dfrac{c_{14}(c_+-2)+2c_+}{2c_{14}(1-c_+)}\mu_1\\\\\dfrac{c_{14}(c_+-2)+2c_+}{2c_{14}(1-c_+)}\mu_2\\\\0\\\\0\end{array}\right),
\left(\begin{array}{c}-2 \mu_1\\\\1\\\\V_3^{1\mu}\\\\-\mu_1\\\\-\mu_2\end{array}\right),
\left(\begin{array}{c}-2 \mu_2\\\\V_2^{1\mu}\\\\1\\\\\mu_2\\\\-\mu_1\end{array}\right),
\left(\begin{array}{c}0\\\\\dfrac{c_+}{1-c_+} \mu_1\\\\-\dfrac{c_+}{1-c_+} \mu_2\\\\1\\\\V_5^{1\mu}\end{array}\right),
\left(\begin{array}{c}0\\\\\dfrac{c_+}{1-c_+} \mu_2\\\\\dfrac{c_+}{1-c_+} \mu_1\\\\V_4^{1\mu}\\\\1\end{array}\right)
\right\}.
\end{align}
\end{widetext}

The propagation eigenstates are
\begingroup
\allowdisplaybreaks
\begin{align}
	H_1&=F+\frac{c_{14}(c_+-2)+2c_+}{2c_{14}(1-c_+)} (\mu_1 \gamma_1+\mu_2\gamma_2),\\
	H_2&=-2\mu_1 F+\gamma_1+V_3^{1\mu} \gamma_2-\mu_1 \phi_{11}-\mu_2 \phi_{12},\\
	H_3&=-2\mu_2 F+V_2^{1\mu} \gamma_1+\gamma_2+\mu_2 \phi_{11}-\mu_1 \phi_{12},\\
	H_4&=\frac{c_+}{1-c_+}(\mu_1 \gamma_1-\mu_2 \gamma_2)+\phi_{11}+V_5^{1\mu} \phi_{12},\\
	H_5&=\frac{c_+}{1-c_+}(\mu_2 \gamma_1+\mu_1 \gamma_2)+V_4^{1\mu} \phi_{11}+\phi_{12}.
\end{align}
\endgroup
We find that the scalar mode and tensor modes both mix with the vector modes but not with each other. Mixing between the two vector modes and the two tensor modes is determined by
constants of first order in $\mu_k$. We cannot determine the values of the
$V_i^{1\mu}$
without going to higher orders in $\mu_k$,\footnote{The approach we use for determining eigenvalues and eigenvectors is essentially perturbation theory that has both degenerate eigenvalues and a nonsymmetric perturbation. At first order in $\mu_k$ we cannot determine the $V_i^{1\mu}$ because the degeneracies are not lifted. At any higher order $n$ we can no longer assume the $n-1$ order contribution to the eigenvectors to be orthogonal. Due to these two issues standard perturbation theory methods fail.} but we also do not need to: they do not affect the eigenvalues and the mode speeds that will lead to observable time lags.
These coefficients amount to a rotation of the two polarizations of equal spin
within the propagation eigenstate to which they dominate: that is, $V_4^{1\mu}, V_5^{1\mu}$ change the relative contribution of $\phi_{11}, \phi_{12}$ in the mostly tensor states, $H_4,H_5$ respectively (and similarly for vector polarizations $\gamma_{1,2}$ within the mostly vector eigenstates $H_2,H_3$). This suggests that the effect of the $V_i^{1\mu}$
is degenerate with the polarization angle,
and would require additional information to be probed.

\subsection{Propagation eigenstate speeds}\label{subsec:propagation_eigenstate_speeds}

To find the propagation eigenstate speeds we set the eigenvalues of $\hat{K}$ to zero. This is equivalent to solving the wave equation or to finding the dispersion relation in the high frequency limit. Similar to the previous section we now expand $\omega$ as $\omega^{0\mu}+\omega^{1\mu}$.
The new zeroth and first orders of the three nondegenerate eigenvalues are:
\begin{align}
E^{0\mu} &\in \bigg\{
\left(\omega^{0\mu}\right)^2-c_S^2 k^2,
\nn\\&\mathrel{\phantom{\in \bigg\{}}
\left(\omega^{0\mu}\right)^2-c_V^2 k^2,
\nn\\&\mathrel{\phantom{\in \bigg\{}}
\left(\omega^{0\mu}\right)^2-c_T^2 k^2
\bigg\},
\\
E^{1\mu} &\in \bigg\{
2\omega^{0\mu} \omega^{1\mu}+d_1 \omega^{0\mu} k \mu_3,
\nn\\&\mathrel{\phantom{\in \bigg\{}}
2\omega^{0\mu} \omega^{1\mu}+\frac{\left(\omega^{0\mu}\right)^3}{k} \mu_3 + a_3 \omega^{0\mu} k \mu_3,
\nn\\&\mathrel{\phantom{\in \bigg\{}}
2\omega^{0\mu} \omega^{1\mu}+b_3 \omega^{0\mu} k \mu_3
\bigg\}.
\end{align}

Notably this
expansion does not change our results. We would arrive at the same speeds even if we had expanded $\omega$ at an earlier step. Setting the zeroth-order eigenvalues to zero we, unsurprisingly, arrive at $\omega^{0\mu}/k=c_S$, etc.
Inserting the zeroth-order solutions into the first order yields
\begin{align}
0&=2\omega^{1\mu}+d_1 k \mu_3,\\
0&=2\omega^{1\mu}+c_V^2 k \mu_3+a_3 k \mu_3,\\
0&=2\omega^{1\mu}+b_3 k \mu_3.
\end{align}
Rearranging, we find
\begin{align}
\frac{\omega^{1\mu}}{k}&=-\frac{d_1}{2} d_1 \mu_3,\\
\frac{\omega^{1\mu}}{k}&=-\frac{a_3+c_V^2}{2} \mu_3,\\
\frac{\omega^{1\mu}}{k}&=-\frac{b_3}{2} \mu_3.
\end{align}

We can now combine zeroth and first orders to obtain the modified speeds $c_S'$, $c_V'$, and $c_T'$. We also calculate and insert the three coefficients $d_1/2$, $b_3/2$ and $(a_3+c_V^2)/2$:
\begin{align}
c_S'&=c_S-\frac{c_{123}}{(1-c_+)(2+c_++3c_2)}\
\nn\\&\mathrel{\phantom{=c_S-}}
\times\left[
c_++2c_2
-\frac{2(1+c_2)^2}{c_{123}}
+ \frac{2}{c_{14}}
\right] \mu_3,\\
c_V'&=c_V-\frac{c_+(c_++2c_4)-2c_4}{2c_{14} (1-c_+)} \mu_3,\\
c_T'&=c_T-\frac{c_+}{1-c_+} \mu_3\label{eq:linear_speed_correction}.
\end{align}

At this point we take stock of how we can apply our results to observations of lensed GW signals. We are primarily interested in how the additional modes arising in Einstein-aether theory affect the speed of tensor waves. This means we assume that scalar and vector modes are not emitted and that we cannot measure $c_S'$ and $c_V'$.

We find that to first order in $\mu_k$ there is only one modified tensor mode speed $c_T'$. There is no birefringence --- that would require two different speeds $c_{T,1}'$ and $c_{T,2}'$. Additionally $c_T'$ depends only on $c_+$ and $\mu_3$:
\begin{align}
	c_T'=\sqrt{\frac{1}{1-c_+}}-\frac{c_+}{1-c_+} \mu_3.
\end{align}
This means that even if we were to observe one or more lensed signal (and model their $\mu_3$) we can at best constrain $c_+$. Since $c_+$ can already be constrained by observing unlensed signals this is not particularly useful. In light of this we now turn to higher orders in $\mu_k$
, with the hope that either birefringence arises or that the higher-order corrections to $c_T$ depend on parameters other than $c_+$.

\section{Higher orders in \texorpdfstring{\textit{\textmu\textsubscript{k}}}{μₖ}}\label{sec:higher_orders}

In this section we investigate higher orders in $\mu_k$. To that end we calculate the full kinetic matrix in Sec.~\ref{subsec:full_kinetic_matrix}. In Sec.~\ref{subsec:second_order}, we use this result to determine the modified mode speeds up to second order in $\mu_k$. After this example calculation, we obtain and discuss the modified mode speeds up to fifth order in $\mu_k$ in Sec.~\ref{subsec:fifth_order}.

\subsection{Full kinetic matrix}\label{subsec:full_kinetic_matrix}

At higher orders in $\mu_k$ many intermediate results become quite complex. We therefore perform most of the remaining calculations with
Mathematica and do not include all of these intermediate results. Some Mathematica-related details of the calculations are discussed in Appendix~\ref{app:mathematica_details}.

We first calculate the full kinetic matrix to all orders in $\mu_k$. To make calculations simpler, we will assume propagation in the $z$ direction from the beginning. Essentially, this will mean that partial derivatives in the $x$ and $y$ directions vanish: $\partial_1=\partial_2=0$. As a consequence, the operator $P_{ij}$
only has the entries $P_{11}=P_{22}=\Delta=\partial_3^2$. The nonvanishing entries of $h_{ij}$ then take the following form:
\begin{align}
h_{01}&=\gamma_1,&
h_{11}&=\frac{F}{2}+\phi_{11},&
h_{12}&=\phi_{12},\\
h_{02}&=\gamma_2,&
h_{22}&=\frac{F}{2}-\phi_{11},&
h_{33}&=\Delta \phi.
\end{align}

We can now calculate the full kinetic matrix. As usual, the first step is the constraint equation, now in its exact form. It separates into parts of zeroth and first order in the perturbation:
\begin{align}
\mu_0&=\sqrt{1+\mu_1^2+\mu_2^2+\mu_3^2},\\
w_0&=
-\frac{\mu_0}{2} h_{00}
-\frac{\mu_1^2+\mu_2^2}{4\mu_0} F
-(\gamma_1 \mu_1+\gamma_2 \mu_2)
\nn\\&\mathrel{\phantom{=}}
-\frac{1}{\mu_0}(v_1 \mu_1+v_2\mu_2)
-\frac{\mu_3^2}{2\mu_0} \Delta \phi
\nn\\&\mathrel{\phantom{=}}
-\frac{\mu_1^2-\mu_2^2}{2\mu_0} \phi_{11}
-\frac{\mu_1 \mu_2}{\mu_0} \phi_{12}.
\end{align}

With this, we now calculate the linearized derivatives, $M$ tensor, aether current, aether stress-energy tensor, $\overline{\text{\AE}^{[0} u^{i]}}$ and effective tensor, as done previously. We do not include their exact forms for brevity. Since we assume propagation in the $z$ direction the SVT decomposition is now simple:
\begin{align}
X_{10}^T&=X_{10},&
X_{20}^T&=X_{20},\\
X_{30}^L&=X_{30},&
X_{01}^T&=X_{01},\\
X_{02}^T&=X_{02},&
X_{03}^L&=X_{03},\\
X_{11}^{TTL}&=\frac{X_{11}-X_{22}}{2},&
X_{12}^{TTL}&=X_{12},\\
X_{11}^{TTF}&=\frac{X_{11}+X_{22}}{2},&
X_{13}^{LTL}&=X_{13},\\
X_{23}^{LTL}&=X_{23},&
X_{33}^{LTF}&=X_{33},
\end{align}
where $TTL$ means the transverse traceless part, $TTF$ means the transverse traceful part, $LTL$ means the longitudinal traceless part, and $LTF$ means the longitudinal traceful part.

We use $X_{00}$, $X_{i0}^T$, and $X_{i0}^L$ to constrain $h_{00}$, $v_1$, $v_2$, and $\phi$. We then apply this to the five equations we need for the kinetic matrix, already properly normalized:
\begin{align}
-\frac{4c_{123}c_{14} X_{11}^{TTF}}{(1-c_+)(2+c_++3c_2)}&=0,\label{eq:X11TTFnorm}\\
-\frac{c_+ X_{01}^T}{(1-c_+)c_{14}}&=0,\\
-\frac{c_+ X_{02}^T}{(1-c_+)c_{14}}&=0,\\
-\frac{2 X_{11}^{TTL}}{1-c_+\mu_0^2}&=0,\\
-\frac{2 X_{12}^{TTL}}{1-c_+\mu_0^2}&=0.\label{eq:X12TTLnorm}
\end{align}

The prefactors of the $F$ terms of the left-hand sides of Eqs.~\eqref{eq:X11TTFnorm} through~\eqref{eq:X12TTLnorm} give us the first row of the kinetic matrix. The $\gamma_1$ terms give us the second row, and so on, until we have obtained the full kinetic matrix.
The result is complex enough that we opted not to calculate the exact eigenvalues and mode speeds. Still, we can now easily obtain the entries of the kinetic matrix and the mode speeds to higher orders in $\mu_k$.

\subsection{Second order in \texorpdfstring{\textit{\textmu\textsubscript{k}}}{μₖ}}\label{subsec:second_order}

As an example of how we obtain the higher-order mode speed corrections we now explicitly calculate the second-order eigenvalues. The second order of the equation defining the eigenvalues and eigenvectors is:
\begingroup
\allowdisplaybreaks
\begin{align}
&
E^{0\mu}\vec{V}^{2\mu}
+E^{1\mu}\vec{V}^{1\mu}
+E^{2\mu}\vec{V}^{0\mu}
\nn\\
&=
\hat{K}^{0\mu}\vec{V}^{2\mu}
+\hat{K}^{1\mu}\vec{V}^{1\mu}
+\hat{K}^{2\mu}\vec{V}^{0\mu}.
\end{align}

The only entries in $\hat{K}^{2\mu}$ that are relevant to us are the fourth and fifth diagonal entries, both of which are equal:
\begin{align}
K_{44}^{2\mu}=K_{55}^{2\mu}&=-(2-c_+)c_+ c_T^4 k^2 \mu_3^2
\nn\\&\mathrel{\phantom{=}}
-\frac{1}{2}\left(2c_+-2c_4-c_+c_-\right)c_T^4 k^2 \left(\mu_1^2+\mu_2^2\right).\label{eq:K44}
\end{align}
\endgroup

\begin{widetext}
We use this and insert $E^{0\mu}$, $E^{1\mu}$, $\vec{V}^{0\mu}$, and $\vec{V}^{1\mu}$ for the first modified tensor polarization.

\begingroup
\allowdisplaybreaks
\begin{align}
	&
	\left(\omega^2-c_T^2 k^2\right)
	\left(\begin{array}{c}
		V_1\\[2mm]V_2\\[2mm]V_3\\[2mm]V_4\\[2mm]V_5
	\end{array}\right)^{2\mu}
	+\frac{2c_+}{1-c_+} \omega k \mu_3
	\left(\begin{array}{c}
		0\\[2mm]\dfrac{c_+}{1-c_+}\mu_1\\[2mm]-\dfrac{c_+}{1-c_+}\mu_2\\[2mm]0\\[2mm]V_5^{1\mu}
	\end{array}\right)
	+E^{2\mu}
	\left(\begin{array}{c}
		0\\\\0\\\\0\\\\1\\\\0
	\end{array}\right)
\nn\\[5mm]=&
 \left(\begin{array}{ccccccccc}
		K_{SS}&&0&&0&&0&&0\\[2mm]
		0&&K_{VV}&&0&&0&&0\\[2mm]
		0&&0&&K_{VV}&&0&&0\\[2mm]
		0&&0&&0&&K_{TT}&&0\\[2mm]
		0&&0&&0&&0&&K_{TT}
	\end{array}\right)^{0\mu}
	\left(\begin{array}{c}
		V_1\\[2mm]V_2\\[2mm]V_3\\[2mm]V_4\\[2mm]V_5
	\end{array}\right)^{2\mu}
	\nn\\[5mm]&
	+
	\left(\begin{array}{ccccccccc}
		K_{SS}^{1\mu}&&K_{SV} \mu_1&&K_{SV} \mu_2&&0&&0\\[2mm]
		K_{VS} \mu_1&&K_{VV}^{1\mu}&&0&&K_{VT} \mu_1 &&K_{VT} \mu_2\\[2mm]
		K_{VS} \mu_2&&0&&K_{VV}^{1\mu}&&-K_{VT} \mu_2 &&K_{VT} \mu_1\\[2mm]
		0&&K_{TV} \mu_1&&-K_{TV}\mu_2&&K_{TT}^{1\mu}&&0\\[2mm]
		0&&K_{TV}\mu_2&&K_{TV} \mu_1&&0&&K_{TT}^{1\mu}
	\end{array}\right)
	\left(\begin{array}{c}
		0\\[2mm]\dfrac{c_+}{1-c_+}\mu_1\\[2mm]-\dfrac{c_+}{1-c_+}\mu_2\\[2mm]0\\[2mm]V_5^{1\mu}
	\end{array}\right)
	+
	\left(\begin{array}{ccccc}
		\ddots&&\vdots&&\vdots\\[2mm]
		\cdots&&K_{44}&&\vdots\\[2mm]
		\cdots&&\cdots&&K_{55}
	\end{array}\right)^{2\mu}
	\left(\begin{array}{c}
		0\\[2mm]0\\[2mm]0\\[2mm]1\\[2mm]0
	\end{array}\right).
\end{align}
\endgroup
\end{widetext}

Since we are only interested in the eigenvalues and speeds, we focus on
the fourth row:
\begin{align}
&
\left(\omega^2-c_T^2 k^2\right)V_4^{2\mu}+E^{2\mu}
\nn\\=&
K_{TT}^{0\mu} V_4^{2\mu}+\frac{K_{TV} c_+}{1-c_+}\left(\mu_1^2+\mu_2^2\right)+K_{44}^{2\mu}.
\end{align}
The first terms on both sides are identical:
\begin{align}
	E^{2\mu}&=\frac{K_{TV} c_+}{1-c_+}\left(\mu_1^2+\mu_2^2\right)+K_{44}^{2\mu}.
\end{align}
We insert $K_{TV}$ and $K_{44}^{2\mu}$ that, remember, are given by Eqs.~\eqref{eq:KTV} and~\eqref{eq:K44}, respectively. We find
\begin{align}
	E^{2\mu}&=\left(-b_1 k^2-b_2 \omega^2\right)\frac{c_+}{1-c_+}\left(\mu_1^2+\mu_2^2\right)
 \nn\\&\mathrel{\phantom{=}}-(2-c_+)c_+ c_T^4 k^2 \mu_3^2
 \nn\\&\mathrel{\phantom{=}}-\frac{1}{2}\left(2c_+-2c_4-c_+c_-\right)c_T^4 k^2 \left(\mu_1^2+\mu_2^2\right).
\end{align}

Finally, we insert $b_1$ and $b_2$:
\begin{align}
	E^{2\mu}
	&=
	\left[-\frac{2c_3+c_+c_-}{2c_+(1-c_+)} k^2+\frac{c_3-c_4}{c_+} \omega^2\right]\frac{c_+}{1-c_+}\left(\mu_1^2+\mu_2^2\right)
 \nn\\&\mathrel{\phantom{=}}-(2-c_+)c_+ c_T^4 k^2 \mu_3^2
 \nn\\&\mathrel{\phantom{=}}-\frac{1}{2}\left(2c_+-2c_4-c_+c_-\right)c_T^4 k^2 \left(\mu_1^2+\mu_2^2\right).
\end{align}

To obtain the second-order speed correction we now expand $\omega$ up to $\omega^{2\mu}$. This again leads to a shuffling of the orders in $\mu_k$:
\begin{align}
E^{2\mu}&=
\left(\omega^{1\mu}\right)^2
+2\omega^{0\mu} \omega^{2\mu}
+\frac{2c_+}{1-c_+} \omega^{1\mu} k \mu_3
\nn\\&\mathrel{\phantom{=}}
+\left[\frac{c_3-c_4}{1-c_+} \left(\omega^{0\mu}\right)^2-\frac{2c_3+c_+c_-}{2(1-c_+)^2} k^2\right]\left(\mu_1^2+\mu_2^2\right)
 \nn\\&\mathrel{\phantom{=}}-(2-c_+)c_+ c_T^4 k^2 \mu_3^2
 \nn\\&\mathrel{\phantom{=}}-\frac{1}{2}\left(2c_+-2c_4-c_+c_-\right)c_T^4 k^2 \left(\mu_1^2+\mu_2^2\right).
\end{align}

We set this expression to zero and insert the values for $\omega^{0\mu}$ and $\omega^{1\mu}$ obtained from the lower orders:
\begin{align}
	0&=
	\frac{c_+^2}{(1-c_+)^2} k^2 \mu_3^2
	+2c_T \omega^{2\mu} k
	-\frac{2c_+}{1-c_+} \frac{c_+}{1-c_+} k^2 \mu_3^2
	\nn\\&\mathrel{\phantom{=}}
	+\left[\frac{c_3-c_4}{1-c_+} c_T^2-\frac{2c_3+c_+c_-}{2(1-c_+)^2} \right]\left(\mu_1^2+\mu_2^2\right)k^2
 \nn\\&\mathrel{\phantom{=}}-(2-c_+)c_+ c_T^4 k^2 \mu_3^2
 \nn\\&\mathrel{\phantom{=}}-\frac{1}{2}\left(2c_+-2c_4-c_+c_-\right)c_T^4 k^2 \left(\mu_1^2+\mu_2^2\right).
\end{align}
We now combine and cancel terms using $1-c_+=c_T^{-2}$:
\begin{align}
0&=
2c_T \omega^{2\mu} k
-2c_+ c_T^4 k^2 \mu_3^2
-c_+ c_T^4 k^2 \left(\mu_1^2+\mu_2^2\right),
\end{align}
and then solve for $\omega^{2\mu}/k$:
\begin{align}
\frac{\omega^{2\mu}}{k}
&=
c_+ c_T^3 \mu_3^2
+\frac{1}{2}c_+ c_T^3 \left(\mu_1^2+\mu_2^2\right).
\end{align}

Repeating the entire calculation for the second modified tensor mode we find that we once again only have one modified mode speed:
\begin{align}
c_T'&=
c_T-c_+ c_T^2 \mu_3
\nn\\&\mathrel{\phantom{=}}
+c_+ c_T^3 \mu_3^2
+\frac{1}{2} c_+ c_T^3 \left(\mu_1^2+\mu_2^2\right).
\label{eq:second_order_speed}
\end{align}

Once again there is no birefringence and the modified speed depends only on $c_+$ and the $\mu_i$. We can now keep repeating the eigenvalue calculation in the same manner for
increasingly higher orders in $\mu_k$.

\subsection{Fifth order in \texorpdfstring{\textit{\textmu\textsubscript{k}}}{μₖ}}
\label{subsec:fifth_order}

For this paper, we calculated the modified tensor speed up to $\mathcal{O}\left(\mu_k^5\right)$:
\begin{align}
c_T' &= c_T \bigg[
1
-c_+ c_T \mu_L
+\frac{1}{2} c_+ c_T^2 \left(2\mu_L^2+\mu_T^2\right)
\nn\\&\mathrel{\phantom{=c_T \bigg[}}
-\frac{1}{2} c_+ c_T^3 \left(1-c_+\right) \mu_L \left(\mu_L^2+\mu_T^2\right)
\nn\\&\mathrel{\phantom{=c_T \bigg[}}
+\frac{1}{8} c_+^2 c_T^4 \left(8\mu_L^4+12\mu_L^2\mu_T^2+3\mu_T^4\right)
\nn\\&\mathrel{\phantom{=c_T \bigg[}}
+\frac{1}{8} c_+ c_T^5 \left(1-6c_+-3c_+^2\right) \mu_L \left(\mu_L^2+\mu_T^2\right)^2
\bigg].
\label{eq:fifth_order_speed}
\end{align}
Here we switched notation to longitudinal ($\mu_L=\mu_3$) and transverse ($\mu_T=\sqrt{\mu_1^2+\mu_2^2}$) aether components. We again find
the absence of birefringence and a modified speed depending only on $c_+$. We suspect that this trend will continue at higher orders and that, unfortunately, we can only constrain $c_+$ with our methods.

\begin{figure}[!th]
\centering
\includegraphics[width=\linewidth]{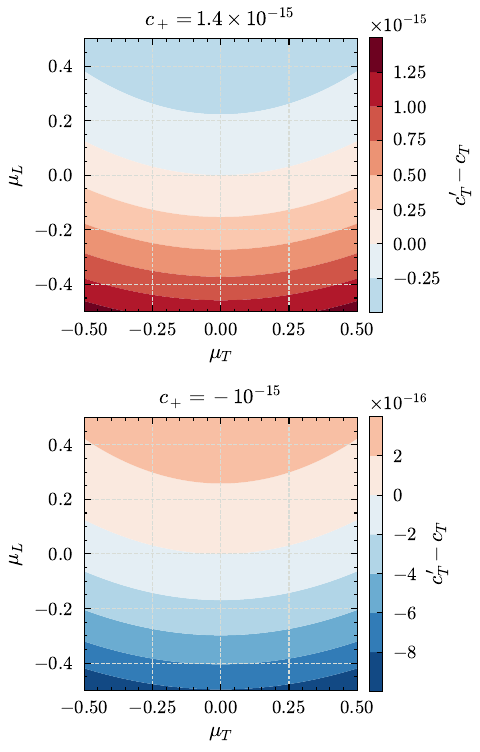}
\caption[Deviation of the Fifth Order Modified Tensor Mode Speed From the Background Value]{Deviation of the fifth-order modified tensor mode speed $c_T'$ from the background value $c_T$ as a function of the longitudinal aether component $\mu_L$ and transverse aether component $\mu_T$. Top: Upper limit of $c_+$ permitted by observations. Bottom: Lower limit of $c_+$. A faster speed compared to the background value is indicated by reds, and a slower speed is indicated by blues. Higher values of $\mu_L$ slow tensor waves down for the upper limit of $c_+$ and speed them up for the lower limit. Higher magnitudes of $\mu_T$ increase speed for the upper limit and decrease it for the lower limit.}
\label{fig:modified_tensor_mode_speed}
\end{figure}

To make the results more intuitive, Fig.~\ref{fig:modified_tensor_mode_speed} shows the effect of the aether on the tensor mode speed by plotting $c_T'-c_T$ (up to fifth order in $\mu_k$) as a function of $\mu_L$, $\mu_T$ and $c_+$. The magnitudes $\abs{\mu_T}$ and $\abs{\mu_L}$ should be smaller than $1$, so we (somewhat arbitrarily) let them run from $-1/2$ to $1/2$. For $c_+$, we recall that observations constrain $\Delta c_T= c_T-1$ to between $-5\times10^{-16}$~\cite{Elliott:2005va} and $7\times10^{-16}$~\cite{Oost:2018tcv}. We can express $c_+$ as a function of $\Delta c_T$:
\begin{align}
c_+=1-\frac{1}{(1+\Delta c_T)^2}\approx 2 \Delta c_T.
\end{align}
Therefore, the two extreme values of $c_+$ permitted by observations are $-10^{-15}$ and $1.4\times 10^{-15}$. We use these values for Fig.~\ref{fig:modified_tensor_mode_speed}.

We find that for positive $c_+$ the longitudinal component of the aether, $\mu_L$, slows down tensor waves.
Larger values of $\mu_L$ correspond to a stronger slowdown, while negative values of $\mu_L$ correspond to negative slowdown, that is, to a  speedup. For negative values of $c_+$ the longitudinal component of the aether instead speeds tensor waves up, acting as a sort of tailwind. The transverse component of the aether, $\mu_T$, has the same effect independent of its sign: it increases $c_T'$ for positive values of $c_+$ and  decreases it for negative values of $c_+$.

\begin{figure}[t]
	\centering
	\includegraphics[width=\linewidth]{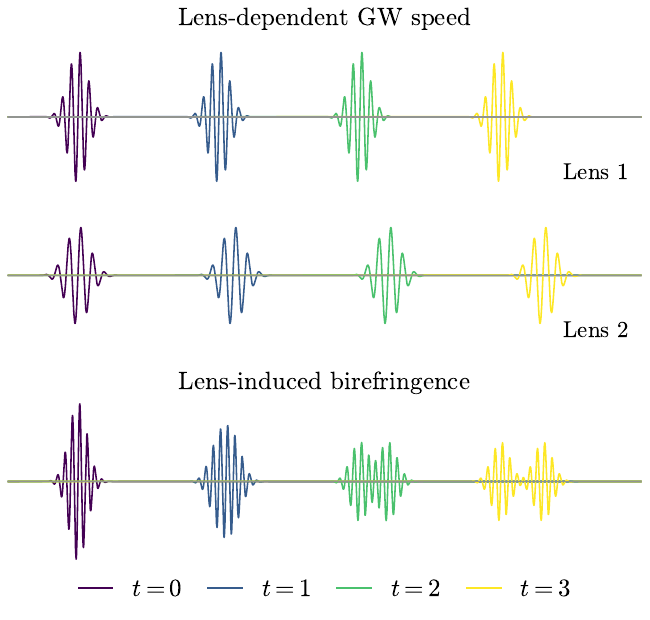}
	\caption[Differences Between Lens-Induced Effects]{Differences between lens-induced effects in different theories.
    The propagation of a wave packet from left to right is shown. Time is measured in arbitrary units. Top and middle:
    Lens-dependent GW speed as observed in Einstein-aether theory. There is no difference in speed between the two tensor modes. There is no birefringence, any one signal does not get scrambled. However, signals lensed by different lenses propagate with different speeds.
    Bottom:
    Lens-induced birefringence as observed in Horndeski's theory. The modified cross and plus mode propagate with different speeds. Birefringence scrambles the signal.
    }
	\label{fig:scenarios}
\end{figure}

The contrast between Einstein-aether theory and Horndeski's theory, as investigated in Ref.~\cite{Ezquiaga:2020dao}, is notable:
\begin{enumerate}
	\item In Einstein-aether theory
    there is only one modified speed $c_T'$ and no birefringence. However, different lensed signals have different speeds. We refer to this as \textit{lens-dependent GW speed}. Observing lensed signals allows us to constrain the parameter combinations that appear in the singular modified speed $c_T'$.
    \item In Horndeski's theory there are two different modified mode speeds $c_{T,1}'$ and $c_{T,2}'$. In this case we have \textit{lens-induced birefringence}. We can measure the delay between the two polarizations of a signal and constrain parameter combinations that appear in $c_{T,1}'-c_{T,2}'$. Note that both speeds depend on the properties of the lens and differently lensed signals may exhibit different time delays.
 \end{enumerate}
Testing these situations requires comparing the arrival time of different signals. This happens because the total travel time is subject to very large uncertainties, e.\,g.,~from gravitational potentials of intervening galaxies~\cite{Boran:2017rdn}.
Testing lens-dependent GW speed therefore requires an electromagnetic or neutrino counterpart~\cite{LIGOScientific:2017zic,Guepin:2022qpl}. This reduces the number of testable events and their redshift, limiting the chances of a close lens-source alignment. However, in the case of lens-induced birefringence, each GW polarization acts as an independent messenger. This allows all GW events to be considered, regardless of nongravitational counterparts~\cite{Goyal:2023uvm}.
Figure~\ref{fig:scenarios} showcases the differences between the two lens-induced effects.

Regarding observations, the main signature is
the dependence of $c_T'$ on $\mu_L$, $\mu_T$, and $c_+$.
To test it one needs to complete the following steps:
\begin{enumerate}
	\item measure $c_T'$ for a lensed signal;
	\item model $\mu_L$ and $\mu_T$ for the signal;
	\item calculate $c_+$ by using Eq.~\eqref{eq:fifth_order_speed};
	\item compare with other constraints imposed on $c_+$.
\end{enumerate}
With this result we now move to conclude this paper and to give an outlook to the future.

\section{Summary and outlook}\label{sec:summary_outlook}

In this paper, we studied GW lensing in Einstein-aether theory in the eikonal, i.e.,~high frequency limit. We discussed Einstein-aether theory with its unit timelike vector field, the aether. We briefly summarized previous results on the propagation of Einstein-aether excitations over Minkowski spacetime with a timelike aether field. Solving the linearized Einstein-aether field equations shows the existence of three new propagating modes. There is the scalar mode $F$ that contracts and expands an object both longitudinally and transversally and two vector modes $\gamma_1$ and $\gamma_2$.
The propagation speeds of all fluctuations, including the standard GW polarizations, are dependent on the theory parameters.

The novelty of our analysis is to combine Einstein-aether theory and GW lensing, that is, describe the evolution of fluctuations in a nonhomogeneous background (e.g., induced by a gravitational lens). We found that approximations are necessary to allow analytical calculations.
We made three such approximations: small GW amplitudes, small background aether inhomogeneities
and high GW frequency. This allowed us to bring the Einstein-aether field equations into the form of a matrix equation containing the kinetic matrix $\hat{K}$. The propagation eigenstate framework~\cite{Ezquiaga:2020dao} (see also Ref.~\cite{Dalang:2020eaj}) allowed us to use eigenvalues and eigenvectors of $\hat{K}$ to study the mixing of modes and the resulting modified mode speeds.

Implementing the small background aether inhomogeneities approximation proved challenging. By expanding to only the first order of the spatial components of the background aether $\mu_k$, calculations were simplified considerably.
We expect this to be valid whenever the gravitational potentials are weak, preventing only the study of GW propagation around black holes and neutron stars.
Note that the assumption of weak gravitational fields does not imply weak lensing: our results are valid for sizable angular deflections, as long as gravitational fields remain small (i.e.,~ the impact parameter is much larger than the gravitational radius of the lens). The geometric optics (eikonal) assumption only precludes us from studying the effect of stellar-mass objects for ground detectors and halos well below subgalactic scales ($M \lesssim 10^9 M_{\odot}$) for space-borne interferometers~\cite{Tambalo:2022wlm}. Typical macroscopic lenses (e.g.~galaxies and clusters~\cite{Shajib:2022con}) are therefore well described by our formalism.

Background aether inhomogeneities introduce a linear correction to the GW speed [Eq.~\eqref{eq:linear_speed_correction}]. Two features of this correction stand out:
\begin{enumerate}
    \item It is common to both GW polarizations. Therefore, Einstein-aether theory predicts no birefringence, and deviations on the GW speed can only be tested by multimessenger observations, using the arrival time relative to another signal (e.g., an electromagnetic counterpart).
    \item It only depends on $c_+$, the same parameter that controls the speed of GWs on a homogeneous background. Therefore, existing constraints limit the magnitude of lensing-induced corrections and their potential to test additional sectors of Einstein-aether theory.
\end{enumerate}
The situation is substantially different from scalar-tensor gravity, where the existence of birefringence~\cite{Ezquiaga:2020dao} allows stringent tests using catalogs of black hole mergers~\cite{Goyal:2023uvm}.
We verified that these results hold up to fifth order in the background aether inhomogeneities $\mathcal{O}\left(\mu_k^5\right)$, plausibly discarding the possibility that they are due to our approximations. This difference may be due to the lack of coupling between tensor and scalar perturbations mediated by nonhomogeneous aether configurations, at leading order of the high-frequency expansion. In Horndeski, this coupling is mediated by nonlinear derivative interactions $\mathcal{L}_H\supset R(\partial\phi)^2, (\Box\phi)^2,\dots$, which are absent in Einstein-aether theory.

There are many aspects in which GW lensing in Einstein-aether theory should be further studied. An important next step is to consider the next orders in the frequency expansion: corrections $\sim f^0$ will describe the amplitude of the propagating degrees of freedom. At this order, it is possible that gravitational lenses introduce observable effects, in the form of amplitude oscillations~\cite{Max:2017flc}, but between tensor, vector, and scalar polarizations. These effects can be tested by the presence of multiple polarizations at the detector~\cite{Takeda:2021hgo}. Oscillations between states with different helicities are forbidden by Lorentz invariance on a symmetric background and can only be mediated by inhomogeneities in the aether field.
An additional order in the expansion, $\sim f^{-2}$, describes dispersive effects in GW propagation~\cite{Dalang:2021qhu,Menadeo:2024uoq}. GW dispersion affects the waveform, and can thus be tested on all signals, without an electromagnetic counterpart~\cite{LIGOScientific:2021sio}. Dispersion is a distinct signature of massive gravity~\cite{Will:1997bb,deRham:2016nuf} (it is also predicted in GR, but only by strong gravitational fields, i.e. lensing by black holes, leading also to multiple images~\cite{Andersson:2020gsj,Oancea:2022szu,Oancea:2023hgu}). However, GW dispersion
and additional polarizations would be a clear signature of lens-induced and beyond-GR effects.

Further studies should overcome the perturbative expansion on the spatial components of the aether field $\mu_k\ll 1$. As a first attempt, this sufficed to characterize the lensing effect on the GW speed. An analysis without assuming small inhomogeneities could be done in general or relying on spherically symmetric solutions of the theory~\cite{Eling:2006df}.
Moreover, a nonperturbative analysis will shed light on the many cancellations leading to the properties of modified GW speed: a lack of birefringence in contrast to scalar-tensor theories~\cite{Ezquiaga:2020dao} and the fact that lens-induced corrections only depend on $c_+$, the one-parameter combination responsible for the GW speed on homogeneous backgrounds. Identifying the structural property of the theory responsible for this feature will likely shed light on how birefringence and anomalous propagation may emerge in other classes of deviations from Einstein's theory.

In this paper we used the propagation eigenstate framework to study Einstein-aether theory. However, we can in principle apply it to many other theories beyond GR, with Lorentz-violating scenarios being especially compelling.
Hořava-Lifshitz gravity~\cite{Horava:2009uw} is a theory of quantum gravity built with concepts from condensed matter physics. Just as with Einstein-aether theory, it breaks local Lorentz invariance.
However, while Einstein-aether theory is equally anisotropic at all energy ranges, Hořava-Lifshitz gravity is strongly anisotropic in the quantum regime: at high energies, high frequencies, short wavelengths, and short length scales. At low energies and long length scales, in the regime of classical gravity, the anisotropy is much weaker~\cite{Horava:2009uw}. Blas-Pujolàs-Sibiryakov-Hořava (BPSH) theory~\cite{Blas:2009qj} is another interesting extension. While Hořava-Lifshitz gravity has a static preferred direction, BPSH theory has a dynamic preferred direction described by a vector field, analogous to Einstein-aether theory. Compared to Einstein-aether theory, there is an additional demand on the vector field in BPSH theory: it must be hypersurface orthogonal. This means that it must be possible to split spacetime into infinitely many spacelike, three-dimensional hypersurfaces so that the vector field is always orthogonal to these hypersurfaces. In fact, Einstein-aether theory is the hypersurface orthogonal low energy limit of BPSH theory~\cite{Jacobson:2010mx}.
We also expect that the rich structure of GW lensing can be linked to the dynamics of gravitational potentials in alternative theories: this will extend known relations in isotropic and homogeneous spacetimes~\cite{Saltas:2014dha,Sawicki:2016klv,Amendola:2017orw} and unlock new consistency tests of gravity theories.

Our work is a first step toward understanding GW propagation in Einstein-aether theories with inhomogeneous field configurations, that is, GW lensing.
The leading order results reflect a remarkable simplicity, as lens-induced corrections are common to both $+,\times$ polarizations and depend only on the parameters that modify the GW speed on a spatially homogeneous background.  This feature, likely anchored in the inner structure of the theory, calls out for a simpler explanation.
It is nevertheless plausible that additional effects in the amplitude and phase of GWs will enable novel tests, not only of Einstein-aether, but of Lorentz invariance in general.
Given the vast numbers of signals available to next-generation ground detectors~\cite{Maggiore:2019uih,Evans:2021gyd,Kalogera:2021bya}, space interferometers pushing towards lower frequencies~\cite{LISA:2017pwj,LISACosmologyWorkingGroup:2022jok,Gong:2021gvw}, and growing evidence for
nanohertz GWs from pulsar-timing arrays~\cite{EPTA:2023fyk,NANOGrav:2023gor,Reardon:2023gzh,Xu:2023wog}, further developments in GW lensing will vastly enhance our capacity to test Einstein's theory and its underlying assumptions.

\section*{Acknowledgements}
We thank Guilherme Brando, Sumanta Chakraborty, Lyla Choi, Alessandro Longo, Stefano Savastano, Hiroki Takeda, Giovanni Tambalo, and H\'ector Villarrubia-Rojo for discussions and the anonymous referee for valuable comments.
H.O.S. acknowledges funding from the Deutsche Forschungsgemeinschaft
(DFG)~-~Project No.~386119226.

\appendix

\section{Effective tensor subtleties}\label{app:effective_tensor_subtleties}

Our definition of the effective tensor deviates from the one used by Foster~\cite{Foster:2006az} in one key point: He includes the term $\overline{\text{\AE}^{[i} u^{j]}}$ in the definition of $X^{ij}$. Physically, this ensures the conservation of source terms. We will refer to his definition of the effective tensor as $\tilde{X}_{\alpha\beta}$. Assuming a sufficiently slowly varying background aether $\mu^\alpha$, the two sets of definitions are related as follows:
\begin{align}
\tilde{X}^{00}&=X^{00},\\
\tilde{X}^{0i}&=X^{0i},\\
\tilde{X}^{i0}&=X^{i0},\\
\tilde{X}^{ij}&=X^{ij}+\frac{\mu^i}{2\mu^0} \cdot X^{0j}-\frac{\mu^i}{2\mu^0} \cdot X^{j0}
\nn\\&\mathrel{\phantom{=}}
-\frac{\mu^j}{2\mu^0} \cdot X^{0i}+\frac{\mu^j}{2\mu^0} \cdot X^{i0},\\
X^{ij}&=\tilde{X}^{ij}-\frac{\mu^i}{2\mu^0} \cdot \tilde{X}^{0j}+\frac{\mu^i}{2\mu^0} \cdot \tilde{X}^{j0}
\nn\\&\mathrel{\phantom{=}}
+\frac{\mu^j}{2\mu^0} \cdot \tilde{X}^{0i}-\frac{\mu^j}{2\mu^0} \cdot \tilde{X}^{i0}.
\end{align}

As $X^{\alpha\beta}$ and $\tilde{X}^{\alpha\beta}$ can be written as linear combinations of each other's components, so can the field equations $X^{\alpha\beta}=0$ and $\tilde{X}^{\alpha\beta}=0$. Therefore, both sets of field equations are mathematically equivalent.

Foster's definition works well enough in the exact case, but becomes problematic when combined with series expansions in $\mu^k$ as used in Sec.~\ref{sec:gw_lensing}. $\tilde{X}^{\alpha\beta}$ is subject to the conservation law $\tilde{X}^{\alpha\beta}_{\phantom{\alpha\beta},\beta}=0$, which is crucial for this definition to yield the correct number of wave equations. Similarly important is the condition
\begin{align}
\tilde{X}^{[ij]}=\frac{\mu^j}{\mu^0} \tilde{X}^{[i0]}+\frac{\mu^i}{\mu^0} \tilde{X}^{[0j]},\label{eq:independent_component_condition}
\end{align}
which stems from $\text{\AE}^{[\alpha} u^{\beta]}$ only having three independent components. Now consider the antisymmetric part of the $i$ component of the conservation law
\begin{align}
\partial_0 \tilde{X}^{[i0]}=\partial_j \tilde{X}^{[ij]}.\label{eq:conservation_law}
\end{align}

If we apply $\partial_j$ to Eq.~\eqref{eq:independent_component_condition}, we can combine it with Eq.~\eqref{eq:conservation_law} into a condition that must be upheld if we want to obtain the correct number of wave equations:
\begin{align}
\partial_0 \tilde{X}^{[i0]}=\frac{\mu^j}{\mu^0} \partial_j \tilde{X}^{[i0]}+\frac{\mu^i}{\mu^0} \partial_j \tilde{X}^{[0j]}.\label{eq:redundancy_condition}
\end{align}

If we violate Eq.~\eqref{eq:redundancy_condition} then we lose one of the two important redundancies and will, in the end, obtain superfluous false wave equations. This means that if we consider a series expansion in $\mu^k$, Eq.~\eqref{eq:redundancy_condition} must hold at all orders and all terms on both sides must always be expanded to the same order in $\mu^k$. However, this is clearly impossible: If $\partial_0 \tilde{X}^{[i0]}$ requires $\tilde{X}^{[i0]}$ to be expanded to the $n$th order in $\mu^k$, then $\mu^j \partial_j \tilde{X}^{[i0]}/\mu^0$, which contains the additional factor $\mu^j$, requires $\tilde{X}^{[i0]}$ to be expanded to $(n-1)$-th order in $\mu^k$. As we cannot expand $\tilde{X}^{[i0]}$ to different orders simultaneously, we find that Foster's definition is unsuitable for us and opt to trade the conservation law for mathematical simplicity.

\section{Gauge considerations}
\label{app:gauge_considerations}

To prove that $\gamma=\phi_i=v=0$ is a possible gauge, we consider how a gauge transformation affects perturbations on a Minkowski background:
\begin{align}
h'_{\mu\nu}
&=
h_{\mu\nu}
+\varepsilon_{\mu,\nu}
+\varepsilon_{\nu,\mu}
\label{eq:gauge_transformation_h},\\
w'_{\mu}
&=
w_\mu
-\mu^\alpha \partial_\alpha \varepsilon_\mu
+\varepsilon^\alpha \partial_\alpha \mu_\mu.
\label{eq:gauge_transformation_w}
\end{align}

Note that some authors switch primed and unprimed quantities, or, conversely, switch the sign for the $\varepsilon$ terms. Performing another SVT decomposition we split $\varepsilon_\mu$ into temporal part $\xi_0$, spatial transverse part $\xi_i$, and spatial longitudinal part $\xi_{,i}$. We then find the following transformation rules for the metric perturbation:
\begin{align}
h_{00}&\rightarrow h_{00}+2\dot\xi_0,\\
\gamma_i&\rightarrow \gamma_i+\dot\xi_i,\\
\gamma &\rightarrow \gamma+ \xi_0 + \dot\xi,\\
\phi_{ij}&\rightarrow \phi_{ij},\\
f&\rightarrow f,\\
\phi_i&\rightarrow \phi_i + \xi_i,\\
\phi&\rightarrow\phi + 2\xi.
\end{align}

For the aether perturbation we will assume a more general background $\mu_\alpha$ instead of $\delta^0_\alpha$. We then also need to decompose the spatial background aether $\mu_i$ into transverse part $\nu_i$ and longitudinal part $\nu_{,i}$:
\begin{align}
w_0
&\rightarrow
w_0
-\mu^\alpha \partial_\alpha \xi_0
-\xi_0 \dot\mu_0
+\xi_k \mu_{0,k}
+\xi_{,k} \mu_{0,k},
\\
v_i
&\rightarrow
v_i
-\mu^\alpha \partial_\alpha \xi_i
-\xi_0 \dot\nu_i
+\xi_k \nu_{i,k}
+\xi_{,k} \nu_{i,k},
\\
v
&\rightarrow
v
-\mu^\alpha \partial_\alpha \xi
-\xi_0 \dot\nu
+\xi_k \nu_{,k}
+\xi_{,k} \nu_{,k}.
\end{align}

Whether or not a gauge is possible to reach depends on whether or not $\xi_0$, $\xi_i$, and $\xi$ can be found for it. For the gauge $\gamma=\phi_i=v=0$, $\xi_0$ and $\xi_i$ can be specified while $\xi$ is the solution of a partial differential equation:
\begin{align}
\xi_0&=-\gamma - \dot\xi,\\
\xi_i&=-\phi_i,\\
\dot\xi&=A_k \xi_{,k}+B,\\
A_i&= \frac{\nu_i}{\mu_0+\dot\nu},\\
B&= -\frac{v-\phi_k \nu_{,k}+\gamma \dot\nu}{\mu_0+\dot\nu}.
\end{align}

We are unable to solve this equation, but to enable our gauge transformation we only need to prove the existence of a solution. This existence is guaranteed by the Cauchy--Kowalevski theorem. We will adopt it from a textbook by
Folland~\cite{Folland_1995} to our specific situation:
\begin{itemize}
	\item One single partial differential equation.
	\item $A_i$, $B$ and $\xi$ depend on spacetime coordinates $x^\mu$ only.
	\item $A_i$ and $B$ are defined on all of spacetime (due to $\mu_0\approx 1$ and $\nu\approx 0$ the denominators cause no issues).
\end{itemize}

The theorem then takes the following form:

Suppose that $A_k$ and $B$ are analytic $\mathbb{R}$-valued functions defined on all of spacetime. Then the Cauchy problem $\dot\xi = A_k\xi_{,k}+B$ with initial conditions $\xi(t=0,x^i)=0$ has a unique analytic solution on all of spacetime.

This theorem ensures that our choice of gauge is possible.

\section{Derivation of the effects of Einstein-aether waves}

\label{app:einstein_aether_waves}

We recall our starting points: monochromatic plane waves in the $z$ direction and the metric and aether perturbations $h_{\mu\nu}$ and $w^\mu$:
\begin{align}
&\psi=\hat\psi \exp\left[i \omega \left(\frac{z}{c_p}-t\right)-\frac{i \pi}{2}\right],\\
	&w^\mu=\left(
	\begin{array}{c}
		\dfrac{h_{00}}{2}\\\\v_i
	\end{array}
	\right),\;
	h_{\mu\nu}=
	\left(
	\begin{array}{ccc}
		h_{00}&&\gamma_i\\\\
		\gamma_i&&\phi_{ij}+\dfrac{1}{2}P_{ij}f+\phi_{,ij}
	\end{array}
	\right).
\end{align}

When acting on plane waves, $P_{ij}$ simplifies to
\begin{alignat}{2}
P_{11} \psi &= P_{22} \psi &&= \Delta \psi,\\
P_{i\neq j} \psi &= P_{33} \psi &&= 0.
\end{alignat}

Similarly, $\partial_i \partial_j$ becomes $\delta_{i3} \delta_{j3} \Delta$ if we assume our solution. We combine this with the constraint equation to obtain metric and aether perturbations:
\begin{align}
w^\mu&=\left(\begin{array}{c}
\dfrac{F}{2 c_{14}}\\\\
-\dfrac{\gamma_1}{c_+}\\\\
-\dfrac{\gamma_2}{c_+}\\\\
0
\end{array}\right),\;
\\
h_{\mu\nu}&=
\left(\begin{array}{ccccccc}
\dfrac{F}{c_{14}}&&\gamma_1&&\gamma_2&&0\\\\
\gamma_1&&\phi_{11} + \dfrac{F}{2}&&\phi_{12}&&0\\\\
\gamma_2&&\phi_{12}&&-\phi_{11} + \dfrac{F}{2}&&0\\\\
0&&0&&0&&-\dfrac{1+c_2}{c_{123}} F
\end{array}\right).
\end{align}

While our result for the metric perturbation is correct, it is difficult to describe the effects of GWs in this gauge since the time coordinate is affected. The synchronous gauge, defined by $h_{\alpha 0}=0$, is better suited for this purpose. In theories beyond GR with all mode speeds equal to $1$ the six gauge invariant polarizations can be expressed in terms of the components of $h_{\mu\nu}$~\cite{Poisson_Will_2014}:
\begin{align}
	h_b&= \frac{1}{2}\left(h_{11}+h_{22}\right),&
	h_l&= h_{00} + 2 h_{03} + h_{33},\\
	h_x&= h_{01}+h_{13},&
	h_y&= h_{02}+h_{23},\\
	h_+&= \frac{1}{2}\left(h_{11}-h_{22}\right),&
	h_\times&= h_{12}.
\end{align}

We now need to correct for the fact that Einstein-aether theory has mode speeds different from $1$. To see what new forms these polarizations take we use the (gauge invariant) Ricci tensor. By calculating its components and forming linear combinations, we can find the six polarizations for speeds different from $1$:
\begin{align}
	h_b&= \frac{1}{2} \left(h_{11}+h_{22}\right),&
	h_l&= \frac{1}{c_X^2} h_{00} + \frac{2}{c_X} h_{03} + h_{33},\\
	h_x&= \frac{1}{c_X} h_{01} + h_{13},&
	h_y&= \frac{1}{c_X} h_{02} + h_{23},\\
	h_+&= \frac{1}{2} \left(h_{11}-h_{22}\right),&
	h_\times&= h_{12}.
\end{align}

Here we treat the inverse mode speed $1/c_X$ as an operator, defined as follows:
\begin{align}
	\frac{1}{c_X} \psi &= \psi \sqrt{\frac{\Delta\psi}{\ddot\psi}}.
\end{align}

In other words, the concrete value that $c_X$ takes will depend on the entry of $h_{\mu\nu}$ that it is acting on.

We now want to use the gauge invariant polarizations to express our metric perturbations in the synchronous gauge. The polarizations $h_b$, $h_+$, and $h_\times$ naturally stay the same before and after transformation since we leave $h_{11}$, $h_{12}$, and $h_{22}$ untouched. For the remaining three polarizations the transformed components of $h_{\mu\nu}$, marked with a prime, must be equal to their counterparts from before the transformation:
\begin{align}
\frac{1}{c_X}h'_{01}+h'_{13}&= \frac{1}{c_X}h_{01}+h_{13},\\
\frac{1}{c_X}h'_{02}+h'_{23}&= \frac{1}{c_X}h_{02}+h_{23},\\
\frac{1}{c_X^2}h'_{00} + \frac{2}{c_X} h'_{03} + h'_{33}&= \frac{1}{c_X^2}h_{00} + \frac{2}{c_X} h_{03} + h_{33}.
\end{align}

$h_{\alpha 0}$ vanishes after the transformation while $h_{03}$, $h_{13}$ and $h_{23}$ vanish before
\begin{align}
h'_{13}&= \frac{1}{c_X}h_{01},&
h'_{23}&= \frac{1}{c_X}h_{02},&
h'_{33}&= \frac{1}{c_X^2}h_{00} + h_{33}.
\end{align}

We know that $h_{01}$ and $h_{02}$ are proportional to $\gamma_1$ and $\gamma_2$, so we can replace the $1/c_X$ acting on them with $1/c_V$. Similarly, we know that $h_{00}$ is proportional to $F$. In this case, we replace $c_X$ with $c_S$,
\begin{align}
h'_{13}&= \frac{1}{c_V}h_{01},&
h'_{23}&= \frac{1}{c_V}h_{02},&
h'_{33}&= \frac{1}{c_S^2}h_{00} + h_{33}.
\end{align}

We apply this transformation to the metric perturbation:
\begin{align}
h_{\mu\nu}&=
\left(\begin{array}{ccccccc}
0&&0&&0&&0\\\\
0&&\phi_{11}+\dfrac{F}{2}&&\phi_{12}&&\dfrac{\gamma_1}{c_V}\\\\
0&&\phi_{12}&&-\phi_{11}+\dfrac{F}{2}&&\dfrac{\gamma_2}{c_V}\\\\
0&&\dfrac{\gamma_1}{c_V}&&\dfrac{\gamma_2}{c_V}&&\alpha F
\end{array}\right).
\end{align}

Here we have defined $\alpha= 1/\left(c_S^2 c_{14}\right)-(1+c_2)/c_{123}$. We can also easily express the polarizations in terms of our five modes:
\begin{align}
	h_b&= \frac{F}{2},&
	h_x&= \frac{\gamma_1}{c_V},&
	h_+&= \phi_{11},\\
	h_l&= \alpha F,&
	h_y&= \frac{\gamma_2}{c_V},&
	h_\times&= \phi_{12}.
\end{align}

Since $F$ excites a mixture of breathing and longitudinal mode we study it in a bit more detail. Assuming that only $F$ is excited, the 3D line element takes the following form:
\begin{align}
ds^2&=\left(1+\frac{F}{2}\right)\left(dx^2+dy^2\right)+\left(1+\alpha F\right)dz^2.
\end{align}

We can rewrite this in spherical coordinates with two parameters $\chi$ and $\beta$:
\begin{align}
ds^2&=\chi^2\left(1+\beta \cos 2\theta\right) dr^2,\\
\chi&=\sqrt{\frac{1}{2}\left(2+\frac{F}{2}+\alpha F\right)},\\
\beta&=\frac{F\left(\alpha-1/2\right)}{2+F/2+\alpha F}.
\end{align}
Note that while $\chi$ and $\beta$ on their own can be complex and singular the relevant quantities $\chi^2$ and $\chi^2 \beta$ never are.

We find that the equation for ${\rm d}s^2$ can be approximated as an ellipse equation. Scalar Einstein-aether waves deform a sphere into an ellipsoid. In the $xz$ or $yz$ plane, we have a semimajor axis $a=\chi\sqrt{1+\beta} \, {\rm d}r$ and an area of $\pi \chi^2 \, {\rm d}r^2$. Unlike the tensor or vector modes, the area of the ellipse cross section can change depending on $\chi$ since $F$ excites the breathing mode. The total effect of the scalar wave will depend on the value of $\alpha$, which depends on the coupling parameters of Einstein-aether and can be positive, negative, or zero.

\section{Mathematica details}\label{app:mathematica_details}

Throughout this paper, results were mostly first calculated by hand and then double-checked using Wolfram Mathematica and the Object-oriented General Relativity (OGRe) package for Mathematica~\cite{Shoshany:2021iuc}. This package was also used for the calculations in Sec.~\ref{sec:higher_orders}. For those calculations we kept the high frequency approximation. This means all derivatives yield zero if they do not act on the perturbation. Since no products of perturbations appear, we can then simply treat $\partial_\alpha$ as a multiplier for the sake of speeding up our calculation.

To shorten the result for the full kinetic matrix we used Mathematica's \textsc{Simplify} function and switched back and forth between using $\mu_0$ or inserting $\mu_0=\sqrt{1+\mu_1^2+\mu_2^2+\mu_3^2}$. Still, the smallest size we could obtain in Mathematica was 936.752 bytes. Obtaining the exact mode speeds would entail finding the eigenvalues of the kinetic matrix and setting them to zero. This is equivalent to solving $\det \hat{K}=0$. The large kinetic matrix causes the determinant to blow up to 16 megabytes. This means that finding the exact mode speeds without using more computational power or different algorithms is unfeasible: all attempts ran for many hours before it was decided to manually terminate them.

In Sec.~\ref{subsec:fifth_order} we applied the functions \textsc{Simplify} and \textsc{FullSimplify} to each order of the kinetic matrix to shorten expressions.

\bibliography{gw_lensing}

\end{document}